\documentclass[twocolumn]{aastex63}
\usepackage{graphicx}
\usepackage{dcolumn}
\usepackage{bm}
\usepackage{amsfonts}
\usepackage{color}
\usepackage{newtxtext,amsmath}
\usepackage{appendix}

\begin{document}

\title{
Impact of magnetic field on neutrino-matter interactions in core-collapse supernova
}

\correspondingauthor{Takami Kuroda}
\email{takami.kuroda@aei.mpg.de}

\author[0000-0001-5168-6792]{Takami Kuroda}
\affiliation{Max-Planck-Institut f{\"u}r Gravitationsphysik, Am M{\"u}hlenberg 1, D-14476 Potsdam-Golm, Germany}
\affiliation{Institut f{\"u}r Kernphysik, Technische Universit{\"a}t Darmstadt, Schlossgartenstrasse 2, D-64289 Darmstadt, Germany}


\shorttitle{Impact of magnetic field on neutrino-matter interactions}
\shortauthors{Kuroda}

\begin{abstract}
We explore the impact of magnetic field on neutrino-matter interactions in core-collapse supernova.
We first derive the modified source terms for neutrino-nucleon scattering and neutrino absorption and emission processes in the moment formalism.
Then we perform full relativistic three-dimensional, magnetorotational core-collapse supernova simulations of a 20 $M_\odot$ star with spectral neutrino transport.
Our simulations treat self-consistently the parity violation effects of weak interaction in the presence of external magnetic field.
The result shows a significant global asymmetry, mostly confined in the proto-neutron star, with clearly reflecting the magnetic field structure.
The asymmetric property arises from two factors: the angle between the neutrino flux and magnetic field, and the term, which is parallel to the magnetic field and is also proportional to the deviation of distribution function of neutrinos from thermal equilibrium.
The typical correction value amounts to $\sim1$ \% relative to the total neutrino-matter interaction rate for the magnetic field strength of $\sim 10^{15-16}$~G.
Although these asymmetric properties do not immediately affect the explosion dynamics, our results imply that they would be significant once the neutrinos diffuse out the proto-neutron star core carrying those asymmetries away.
We also show that, during our simulation time of $\sim370$~ms after bounce, our results indicate that the correction value due to the modified inelastic scattering process dominates over that of the modified neutrino absorption and emission process.
\end{abstract}


\section{Introduction}
\label{sec:Introduction}
Massive stars heavier than $\sim8~M_\odot$ terminate their lives with a catastrophic collapse of their central core and subsequent diverse phenomena.
Some of them are accompanied by a huge explosion called core-collapse supernova \citep[see][for recent reviews]{Janka16,BMullerReview16,Radice18}.
There are currently two major explosion mechanisms suggested.
The first one relies on the complex interplay between neutrinos and stellar mantle and is considered to account for CCSNe with a canonical explosion energy of $\sim10^{51}$ ergs$(\equiv1$ Bethe,~1B in short).
The other is called magnetorotational (MR) explosion that takes place if the progenitor star rotates sufficiently fast and is also magnetized \citep{Bisnovatyi-Kogan70,LeBlancWilson70,Meier76,EMuller79}.

The MR explosion is intrinsically an asymmetric phenomenon.
It is characterized by collimated bipolar outflows \citep{arde00,Burrows07,Takiwaki09,Scheidegger10,Winteler12,Sawai16,Obergaulinger20,Bugli20}.
However the recent full three-dimensional (3D) magnetohydrodynamics (MHD) simulations report also less collimated, slightly weaker outflows \citep{Moesta14,Obergaulinger20,KurodaT20}.
The bipolar structure is originated from magnetic field amplification taking place mainly along the rotational axis.
Strongly amplified magnetic fields eventually eject matter toward the rotational axis.
Another aspect of the MR explosion is its relatively high explosion energy compared to the aforementioned neutrino heating-driven explosion.
Depending on the available differential rotational energy of proto-neutron star (PNS), the magnetic field can be very efficiently amplified up to the equipartition level, i.e., roughly the same as the differential rotational energy.
Due to the efficient conversion, typically one order of magnitude larger explosion can be achieved in the previous MHD CCSN models \citep{Burrows07,Takiwaki09,Obergaulinger20,KurodaT20}.

Combination of these two aspects makes the MR explosion as one of plausible mechanisms of a subclass of CCSNe called hypernova (HN), which observationally presents high explosion energy of $\sim$ 10 B \citep[see, e.g.,][and references therein]{Nomoto06} and is also often accompanied by a bipolar explosion \cite[e.g.,][]{Ezzeddine19}.
Although the so-called collapsar model \citep{MacFadyen99} can currently be considered as a promising mechanism of HNe \citep[e.g.,][and references therein]{Proga03,Kumar15}, the MR explosion is also attracting considerable attention  \citep{Metzger11}.
Furthermore, the asymmetric property of MR explosion is also considered as a plausible kick mechanism of compact object.
All CCSNe leave behind a compact object either a neutron star (NS) or a black hole and these compact objects are occasionally observed with a proper motion.
While observational results of the BH kick velocities are still ambiguous (see, e.g., \cite{Repetto12} for high-velocity cases or \cite{Mandel16} for counterarguments), those of NSs show significant velocities of several hundred to $\gtrsim1000$~km~s$^{-1}$ \citep{Hobbs05,Winkler07}.

Such a significant proper motion is likely produced during CCSNe via two possible mechanisms \citep[see, e.g.,][for a review]{Lai01}.
The first one is via asymmetric matter motion and the other is via asymmetric neutrino emission.
Regarding the former mechanism, most of the recent multi-D simulations of CCSNe indicate that the explosion takes place asymmetrically \citep[for both 2D and 3D models, see, e.g.,][]{Bruenn16,takiwaki16,BMuller17,O'Connor18,Pan18,Summa18,Ott18,Nagakura19,Vartanyan19}.
This is also the case for MHD simulations \citep{Sawai08,Scheidegger10,Winteler12,Moesta14,Obergaulinger20,Bugli20,KurodaT20}.
In addition, there are observational supports for these asymmetric CCSN models \citep[e.g.,][]{katsuda18}.
The asymmetric explosion in these numerical simulations is often dominated by low-order spherical harmonics modes.
Then the central NS is considered to recoil from conservation of linear momentum and the estimated kick velocity reaches a few 100 to $\sim1000$~km~s$^{-1}$ \citep{Sawai08,Wongwathanarat13,Janka17,Chan&Muller18,BMuller19,Nakamura19}.

Some progenitor stars, particularly low-mass stars, however, show nearly symmetric ejecta which are inadequate to account for the observed kick velocity \citep{Gessner18}.
In such cases, asymmetric neutrino emission can be another possible origin of the kick velocity.
Analogous to the anisotropic matter ejection, neutrinos exchange momentum with matter and, if the net exchange does not vanish, the PNS is accelerated.
While some previous studies report the acceleration due to the anisotropic neutrino emission is minor and can explain the kick velocity of only up to a few 100~km~s$^{-1}$ \citep{Wongwathanarat13,Tamborra14ApJ,Gessner18,BMuller18}, \cite{Nagakura19} report that its effect is moderate due to self-sustained partial distribution of the electron fraction ($Y_e$).

The net momentum that the central compact object gains depends on the degree of asymmetry of neutrino emission and also on its duration time.
Regarding the former, the lepton-number emission self-sustained asymmetry \citep{Tamborra14ApJ,O'Connor18,Glas19,Vartanyan19} is currently one of the relevant mechanisms to induce the asymmetric neutrino emission.
According to \cite{Tamborra14ApJ}, the LESA can explain the kick velocity of 100-200~km~s$^{-1}$.
The degree of asymmetry of neutrino emission can still increase if we take into account the magnetic field.
In the presence of magnetic field, a global asymmetry could appear as a consequence of parity violation in the neutrino-matter interactions, e.g., neutrino scattering on slightly polarized free nucleons due to magnetic field or Landau quantization of free electrons.

There have been several works studying the parity violation effects of weak interaction in the presence of magnetic in CCSNe.
\cite{Bisnovatyi-Kogan93} investigated the influence of magnetic field on $\beta$-process and derived kick velocity between $\sim100$ and $\sim3000$~km~s$^{-1}$.
\cite{Horowitz98} explored the effect of elastic scattering of neutrinos on slightly polarized free neutrons due to magnetic field and estimated the magnetic field strength $B$ of $\sim10^{14}$~G to account for the kick velocity of $\sim250$~km.
Later \cite{Arras&Lai99} (hereafter AL99) pointed out the importance of inelasticity in the scattering process on free nucleons and also of correct treatment of detailed balance of neutrinos in thermal equilibrium.
Taking these modifications into account, they formulated the scattering cross section as well as the absorption one and showed that a dipole magnetic field of $\gtrsim10^{15-16}$~G is required to generate a kick velocity of a few hundred~km~s$^{-1}$.
\cite{Kotake05} was the first to implement the effect of parity violation by $\beta$-process into the MHD simulation of CCSN and showed the excess/reduction of heating rate of $\sim0.5$ \%.
\cite{Maruyama11,Maruyama12} calculated the neutrino absorption and cross sections in the context of relativistic mean field theory and derived relatively high magnetic field of $\sim10^{17}$~G to explain the kick velocity of $\sim$500-600~km~s$^{-1}$.
In addition to the NS kick, the parity violation effect could also contribute to the pulsar spin evolution \citep{Maruyama14,Suwa14}.

In this paper, we aim to implement  self-consistently the parity violation effects of weak interactions in the presence of magnetic field to full relativistic two-moment (M1) neutrino transport code.
For that purpose, we first rewrite the absorption and cross sections described in AL99 in moment formalism. 
Then we apply those modified source terms to full 3D-GR, MR core-collapse simulations of a 20 $M_\odot$ star with spectral neutrino transport.
We calculate three models: nonrotating nonmagnetized, rotating strongly magnetized, and a supplemental rotating model with ultra-strongly magnetized models.
We also update the neutrino opacities following \cite{Kotake18}.
Our results clearly show fingerprints of parity violation in weak interaction.
Because of the initially dipole like magnetic field employed, the parity violation results in a global asymmetry mostly with respect to the equatorial plane.
Although these asymmetric features do not immediately affect the explosion dynamics, our results imply that they would be significant in the diffusion time scale of neutrinos.
In addition, we also show the importance of the modified inelastic scattering process, which has been often omitted in previous literature, relative to the modified neutrino absorption and emission process.

This paper is organized as follows.
Section~\ref{sec:Basic nu-GRMHD Equations} starts with a concise summary of our GR MHD neutrino transport scheme.
In Section~\ref{sec:Neutrino source terms in the presence of magnetic field}, we shortly explain the scattering and absorption cross sections in the presence of magnetic field and also rewrite them suitable for a moment formalism.
We describe the initial setup of the simulation together with the updated neutrino opacities in Section~\ref{sec:Initial models and neutrino opacities}.
The main results and detailed analysis of the effects of parity violation are presented in Section~\ref{sec:Results}.
We summarize our results and conclude in Section~\ref{sec:Summary and Discussion}.
In Appendix~\ref{app:Lepton number conservation}, we prove that the modified scattering term does not violate the lepton number conservation.
Note that the geometrized unit is used in Section \ref{sec:Neutrino source terms in the presence of magnetic field} unless otherwise stated,
i.e., the speed of light, the gravitational constant, and the Planck constant are set to unity, $c= G = h=1$,
and cgs units are used in Section \ref{sec:Results}.
The metric signature is $(-,+,+,+)$.
Greek indices run from 0 to 3 and Latin indices from 1 to 3, except $\nu$ and $\varepsilon$ that denote neutrino species and energy, respectively.

\section{Basic $\nu$-GRMHD Equations}
\label{sec:Basic nu-GRMHD Equations}
In our full GR radiation-MHD simulations, we solve the evolution equations of metric, MHD, and energy-dependent neutrino radiation.
Each of metric and radiation-MHD parts is solved in an operator-splitting manner, but the system evolves self-consistently as a whole satisfying the Hamiltonian and momentum constraints.
In this paper, we omit to show the evolution equations for the metric and MHD parts, which can be found in our former papers \cite{KurodaT16,KurodaT20}.
We only recapitulate our basic neutrino transport equations below.

We solve the evolution equations for the zeroth $E_{\varepsilon}$ and first order radiation moments $F^{\alpha}_{\varepsilon}$ measured by an Eulerian observer, with $\varepsilon$ representing the neutrino energy measured in the comoving frame.
The evolution equations read \citep{Shibata11}
\begin{eqnarray}
\partial_t \sqrt{\gamma}E_{\varepsilon}+&&\partial_i \sqrt{\gamma}(\alpha F_{\varepsilon}^i-\beta^i E_{\varepsilon})
+\sqrt{\gamma}\alpha \partial_\varepsilon \bigl(\varepsilon \tilde M^\mu_{\varepsilon} n_\mu\bigr)  =\nonumber \\
&&\sqrt{\gamma}(\alpha P^{ij}_{\varepsilon}K_{ij}-F_{\varepsilon}^i\partial_i \alpha-\alpha S_{\varepsilon}^\mu n_\mu)
\label{eq:rad1}
\end{eqnarray}
\noindent
and 
\begin{eqnarray}
&&\partial_t \sqrt{\gamma}{F_{\varepsilon}}_i+\partial_j \sqrt{\gamma}(\alpha 
{P_{\varepsilon}}_i^j-\beta^j {F_{\varepsilon}}_i)
-\sqrt{\gamma}\alpha \partial_\varepsilon\bigl(\varepsilon \tilde M^\mu_{\varepsilon} \gamma_{i\mu}\bigr)=\nonumber \\
&&\sqrt{\gamma}[-E_{\varepsilon}\partial_i\alpha +{F_{\varepsilon}}_j\partial_i \beta^j+(\alpha/2) 
P_{\varepsilon}^{jk}\partial_i \gamma_{jk}+\alpha S^\mu_{\varepsilon} \gamma_{i\mu}].
\label{eq:rad2}
\end{eqnarray}

\noindent

Here, $\alpha$, $\beta^i$, $\gamma^{ij}$, and $K_{ij}$ are the lapse function, shift vector, three metric, and extrinsic curvature, respectively.
$\gamma\equiv{\rm det}(\gamma_{ij})$ is the determinant of the three metric, $P^{ij}$ is the second order radiation moment measured in the Eulerian frame and is given by an analytic closure relation (see \cite{KurodaT16}), and $S^{\mu}_{\varepsilon}$ is the source term for neutrino matter interactions.
According to \cite{Shibata11}, the four source term $S^\mu_\varepsilon$ in the moment formalism can be expressed as
\begin{eqnarray}
S^\mu_\varepsilon=\varepsilon^3\int d\Omega B(\varepsilon,\Omega) (u^\mu+l^\mu),
\label{eq:4sourceterm}
\end{eqnarray}
where $B(\varepsilon,\Omega)$ is the source term for the distribution function of neutrinos $f(\varepsilon,\Omega)$ with $\Omega$ representing the angular dependence.
$l^\mu$ is a unit normal four vector orthogonal to the four velocity $u^\mu$ and is also used to obtain the radiation moments by angular integration of the distribution function as follow
\begin{eqnarray}
(J_\varepsilon,H^\alpha_\varepsilon,L^{\alpha\beta}_\varepsilon)=\varepsilon^3\int d\Omega (1,l^\alpha,l^\alpha l^\beta)f(\varepsilon,\Omega).
\label{eq:JHL}
\end{eqnarray}
$J_\varepsilon$, $H^\alpha_\varepsilon$, and $L^{\alpha\beta}_\varepsilon$ are the zeroth, first, and second order radiation moments measured in the comoving frame, respectively.
In the Doppler and red-shift term, $\tilde M^\mu_{\varepsilon}$ is defined by $\tilde M^\mu_{\varepsilon}\equiv M^{\mu\alpha\beta}_{\varepsilon}\nabla_\beta u_\alpha$, where $M^{\mu\alpha\beta}_{\varepsilon}$ denotes the third rank moment of neutrino distribution function \citep[see][for more detailed expression]{Shibata11}.
In the next section, we explain how $B(\varepsilon,\Omega)$ is modified in the presence of magnetic field and then rewrite it using Eq.~(\ref{eq:4sourceterm}) to obtain the modified four source term which is used in the M1 neutrino transport equations (\ref{eq:rad1})-(\ref{eq:rad2}).

\section{Neutrino source terms in the presence of magnetic field}
\label{sec:Neutrino source terms in the presence of magnetic field}
We begin with a brief description of the modified neutrino source term $B(\varepsilon,\Omega)$ in the presence of magnetic field.
We refer the reader to AL99 for more detailed background.
In the followings, we focus on the impacts of magnetic field on the two dominant neutrino matter interactions inside the PNS: the neutrino-nucleon scattering and the absorption and emission processes on free nucleons.
We now explain each process one by one.

\subsection{Modified neutrino-nucleon scattering rate}
\label{sec:Modified neutrino-nucleon scattering rate}
The neutrino scattering rate on free nucleons $B_{\rm sc}(\varepsilon,\Omega)$ can be written as
\begin{eqnarray}
B_{\rm sc}(\varepsilon,\Omega)=&&\int  d\varepsilon' d\Omega' \frac{d\Gamma}{d\varepsilon' d\Omega'}\nonumber \\
&&\Bigr[ e^{-(\varepsilon-\varepsilon')/T}\bigl(1-f(\varepsilon, \Omega)\bigr)f(\varepsilon', \Omega') \Bigl. \nonumber \\
&&-\bigl(1-f(\varepsilon', \Omega')\bigr)f(\varepsilon, \Omega)
\Bigl. \Bigr].
\label{dfdt_sc}
\end{eqnarray}
Here $T$ is the matter temperature and $d\Gamma/(d\varepsilon' d\Omega')$ is the differential cross section.
Variables with $'$ denote those of outgoing neutrinos.

Modification of the scattering rate due to the magnetic field appears in the differential cross section $d\Gamma/(d\varepsilon' d\Omega')$ in terms of $\mu_B b$.
Here $\mu_B$ and $b$ are the magnetic moment and magnetic field strength, respectively.
We use the value $b=\sqrt{b^\alpha b_\alpha}$ for the magnetic field strength with $b^\alpha$ being the comoving four magnetic field.
The typical value of $|\mu_B b|$ in CCSNe is $\lesssim\mathcal{O}(10^{-2})$ MeV, even when one considers a strongly magnetized neutron star with $b\sim10^{15}$~G \citep[magnetar:][]{Duncan92}.
Such a value is significantly smaller than the matter temperature $T\gtrsim10$ MeV and we can safely neglect the second or higher order terms of $\mu_B b$ in the expansion of $d\Gamma/(d\varepsilon' d\Omega')$.
Then it reads (see AL99)
\begin{eqnarray}
\frac{d\Gamma}{d\varepsilon' d\Omega'}=&&A_0(\varepsilon,\varepsilon',\mu')\nonumber \\
&&+\delta A_+(\varepsilon,\varepsilon',\mu',b)l^\alpha \hat b_\alpha
+\delta A_-(\varepsilon,\varepsilon',\mu',b)l'^\alpha \hat b_\alpha,\nonumber \\
\label{eq:dGamma_nu_e}
\end{eqnarray}
for electron type neutrino scattering $\nu_eN\rightarrow\nu_eN$, and
\begin{eqnarray}
\frac{d\Gamma}{d\varepsilon' d\Omega'}=&&A_0(\varepsilon,\varepsilon',\mu')\nonumber \\
&&+\delta A_-(\varepsilon,\varepsilon',\mu',b)l^\alpha \hat b_\alpha
+\delta A_+(\varepsilon,\varepsilon',\mu',b)l'^\alpha \hat b_\alpha,\nonumber \\
\label{eq:dGamma_bar_nu_e}
\end{eqnarray}
for electron type anti-neutrino scattering $\bar\nu_e N\rightarrow \bar\nu_e N$.

Here $A_0$ and $\delta A_\pm$ are the zeroth and first order terms of Taylor expansion of $d\Gamma/(d\varepsilon' d\Omega')$ about $b$.
$\hat b^\alpha$ is a unit vector parallel to $b^\alpha$.
$\mu'$ is the scattering angle between the incoming and outgoing neutrinos.
From Eqs.~(\ref{eq:dGamma_nu_e})-(\ref{eq:dGamma_bar_nu_e}), one can rewrite the scattering rate $B_{\rm sc}(\varepsilon,\Omega)$ as a sum of normal scattering rate $B_{\rm sc}^{b=0}(\varepsilon,\Omega)$, which is independent on the magnetic field, and correction term $B_{\rm sc}^{b\neq0}(\varepsilon,\Omega)$, which is proportional to $b$, as
\begin{equation}
B_{\rm sc}(\varepsilon,\Omega)=B_{\rm sc}^{b=0}(\varepsilon,\Omega)+B_{\rm sc}^{b\neq0}(\varepsilon,\Omega).
\label{Bsc_b=0,b/=0}
\end{equation}

For reference, the correction terms $\delta A_\pm(\varepsilon,\varepsilon',\mu',b)$ are expressed in cgs units as (AL99)
\begin{flalign}
&\delta A_\pm(\varepsilon,\varepsilon',\mu',b)=\frac{2\varepsilon'^2G_F^2 h_V^N h_A^N m_N^2 \mu_B^N b}{\pi q} \nonumber \\
&\times\frac{1}{\left[\exp(x_0)+1\right]\left[\exp(-x_0-z)+1\right]}\left(1\pm \frac{h_A^N}{h_V^N}\frac{2m_Nq_0}{q^2} \right),\nonumber \\
\label{eq:deltaA}
\end{flalign}
for scattering on $N$, where $N$ takes either $n$(neutron) or $p$(proton).
In Eq.~(\ref{eq:deltaA}), $G_F$ is the Fermi constant with its explicit value being $G_F^2=G_F^2 c/(\hbar c)^4=1.55\times10^{-33}$ [cm$^3$ MeV$^{-2}$ s$^{-1}$].
$h_V^N$ and $h_A^N$ are the neutral nucleon current form factors and we adopt the same values used in \cite{Bruenn85,KurodaT16}.
$\mu_B^N=g_Ne\hbar/(2m_Nc)$ is the nucleon magnetic moment with $g_n=-1.913$ and $g_p=2.793$, $q_0=\varepsilon-\varepsilon'$, $q=\sqrt{\varepsilon^2+\varepsilon'^2-2\varepsilon\varepsilon'\mu'}$, $x_0=\frac{(q_0-q^2/2m_N)^2}{4Tq^2/2m_N}-\frac{\mu_N}{T}$, and $z=q_0/T$.
Hereafter, we omit the arguments $(\varepsilon,\varepsilon',\mu',b)$ in $\delta A_\pm$ for simplicity.

\subsection{Moment formalism of the modified neutrino-nucleon scattering rate}
\label{sec:Moment formalism of the modified neutrino-nucleon scattering rate}
In the moment formalism, the scattering term reads
\begin{eqnarray}
S^\alpha_{\varepsilon,{\rm sc}}
=S^{\alpha,b=0}_{\varepsilon,{\rm sc}}+
S^{\alpha,b\neq0}_{\varepsilon,{\rm sc}},
\label{eq:4sourceterm_sc}
\end{eqnarray}
where
\begin{eqnarray}
S^{\alpha,b=0/b\neq0}_{\varepsilon,{\rm sc}}&=&
\varepsilon^3\int d\Omega B^{b=0/b\neq0}_{\rm sc}(\varepsilon,\Omega) (u^\alpha+l^\alpha).
\label{eq:4sourceterm_sc_b=0_bneq0}
\end{eqnarray}

For the scattering term without the contribution of magnetic field $S^{\alpha,b=0}_{\varepsilon,{\rm sc}}$, we take into account only the isoenergetic scattering.
Here, we note that scattering on free nucleon is not a perfect elastic system, especially when it is not in thermal equilibrium.
However, the inelastic correction term is still minor in CCSNe \citep{Wang20} and, furthermore, to maintain the consistency with our previous studies, we simply use the isoenergetic kernel this time.
It is written as
\begin{eqnarray}
S^{\alpha,b=0}_{\varepsilon,{\rm sc}}&=&\varepsilon^3\int d\Omega B^{b=0}_{\rm sc}(\varepsilon,\Omega) (u^\alpha+l^\alpha)\nonumber \\
&=&-\chi^{\rm iso}_\varepsilon H^\alpha_\varepsilon,
\label{eq:4sourceterm_sc_b=0}
\end{eqnarray}
where $\chi^{\rm iso}_\varepsilon$ is expressed in terms of the isoenergetic scattering kernel \citep[see Appendix 2 in][for the explicit expression]{KurodaT16}.

Now we move on to how the source term $B_{\rm sc}^{b\neq0}(\varepsilon,\Omega)$ can be expressed in the moment formalism.
We note that, in the following, the modified source term $S^{\alpha,b\neq0}_{\varepsilon,{\rm sc}}$ arising from the magnetic field fully considers the inelasticity, which is omitted in the term $S^{\alpha,b=0}_{\varepsilon,{\rm sc}}$.
This is because that the contribution from inelasticity becomes sometime dominant in the final source term $S^{\alpha,b\neq0}_{\varepsilon,{\rm sc}}$.
This is particularly true for neutrinos with energy $\varepsilon\lesssim4T(\sim10-80$ MeV for $T\sim3-20$ MeV) (see the discussion around Eq.~(4.31) in AL99), i.e., most of neutrinos inside the PNS.
Therefore the inelasticity is crucial in the parity-violation term.

In the following, we consider the neutrino-nucleon scattering.
For the anti-neutrino-nucleon scattering, we simply switch $\delta A_+$ and $\delta A_-$ in the equations below.
The first term in the right hand side of Eq.~(\ref{eq:4sourceterm_sc_b=0_bneq0}), which is parallel to $u^\alpha$, corresponds to the zeroth order moment of the source term.
Introducing the following same notations used in AL99 (see their Eqs.~4.14-4.15)
\begin{eqnarray}
C(\varepsilon,\varepsilon')=e^{-q_0/T}(1-f_\varepsilon^{\rm eq})+f_\varepsilon^{\rm eq}
\end{eqnarray}
and
\begin{eqnarray}
D(\varepsilon,\varepsilon')=-\left(e^{-q_0/T}f_{\varepsilon'}^{\rm eq}+1-f_{\varepsilon'}^{\rm eq}\right),
\end{eqnarray}
where $f^{\rm eq}_{\varepsilon}=1/(1+\exp((\varepsilon-\mu_\nu)/T))$ represents the Fermi distribution function of neutrino with $\mu_\nu$ being the chemical potential, the zeroth order term becomes
\begin{flalign}
\varepsilon^3 &\int d\Omega B^{b\neq0}_{\rm sc}(\varepsilon,\Omega)u^\alpha =   \left[H_\varepsilon^\beta a_{1,\beta}(\varepsilon)+4\pi\varepsilon^3c_0(\varepsilon) \right]u^\alpha.
\label{eq:0thangularmoment_sc}
\end{flalign}
Here, we define
\begin{flalign}
a_{1,\alpha}(\varepsilon)=&  2\pi\hat b_\alpha \int d\varepsilon' \varepsilon'^2 D(\varepsilon,\varepsilon') \int d\mu' ( \delta A_{+}+\mu'\delta A_{-}  )
\label{eq:0thangularmoment_sc_a1}
\end{flalign}
and
\begin{flalign}
c_0(\varepsilon)=&  2\pi \hat b_\alpha \int d\varepsilon' \varepsilon'^2 C(\varepsilon,\varepsilon') \frac{H_{\varepsilon'}^\alpha}{4\pi\varepsilon'^3}\int d\mu' ( \mu'\delta A_{+}+\delta A_{-}  ).
\label{eq:0thangularmoment_sc_c0}
\end{flalign}
To derive Eq.~(\ref{eq:0thangularmoment_sc}), we approximate the neutrino distribution function after scattering into isotropic and nonisotropic parts as
\begin{flalign}
f(\varepsilon',\Omega')\approx f^0_{\varepsilon'}+f^{1,\alpha}_{\varepsilon'} l_\alpha',
\label{eq:f_after_sc}
\end{flalign}
where $f^0_{\varepsilon'}$ and $f^{1,\alpha}_{\varepsilon'}$ do not have angle dependency. They are related to the zeroth and first order radiation moments in comoving frame as
\begin{flalign}
f^0_{\varepsilon}=& \frac{J_\varepsilon}{4\pi\varepsilon^3} \label{eq:J_after_sc}\\
f^{1,\alpha}_{\varepsilon}=&\frac{3H_\varepsilon^\alpha}{4\pi\varepsilon^3}.
\label{eq:H_after_sc}
\end{flalign}

Similarly, the first order moment of the source term can also be expressed in terms of the radiation moments as
\begin{eqnarray}
\varepsilon^3 \int d\Omega B^{b\neq0}_{\rm sc}(\varepsilon,\Omega) l^\alpha=   \left[ \tilde L_\varepsilon^{\alpha\beta} a_{1,\beta}(\varepsilon) +\frac{4\pi\varepsilon^3}{3} h^{\alpha\beta} c_{1,\beta}(\varepsilon) \right],\nonumber \\
\label{eq:1stangularmoment_sc}
\end{eqnarray}
where
\begin{eqnarray}
c_{1,\alpha}(\varepsilon)= &&2\pi\hat b_\alpha \int d\varepsilon' \varepsilon'^2 C(\varepsilon,\varepsilon') (J_{\varepsilon'}-J_{\varepsilon'}^{\rm eq})/(4\pi\varepsilon'^3)\nonumber \\
&&\times\int d\mu' ( \delta A_{+}+\mu'\delta A_{-}  ).
\label{eq:1stangularmoment_sc_c1}
\end{eqnarray}
In the equation,
\begin{flalign}
\tilde L_\varepsilon^{\alpha\beta}= L_\varepsilon^{\alpha\beta}- \frac{1}{3}h^{\alpha\beta} J_\varepsilon^{\rm eq},\\
\label{eq:Ltilde}
h_{\alpha\beta}=g_{\alpha\beta}+u_\alpha u_\beta,
\end{flalign}
and
\begin{flalign}
J_\varepsilon^{\rm eq}=4\pi\varepsilon^3f^{\rm eq}_{\varepsilon}.
\end{flalign}

Consequently, the summation of Eqs.~(\ref{eq:0thangularmoment_sc}) and (\ref{eq:1stangularmoment_sc}) yields the final expression of the source term Eq.~(\ref{eq:4sourceterm_sc_b=0_bneq0}) expressed as
\begin{flalign}
S^{\alpha,b\neq0}_{\varepsilon,\rm sc}=&\left[H_\varepsilon^\beta u^\alpha +\tilde L_\varepsilon^{\alpha\beta}  \right] a_{1,\beta}(\varepsilon)\nonumber \\
&+4\pi\varepsilon^3 u^\alpha c_0(\varepsilon) \nonumber \\
&+\frac{4\pi\varepsilon^3}{3} h^{\alpha\beta} c_{1,\beta}(\varepsilon).
\label{eq:Final4sourceterm_sc_bneq0}
\end{flalign}
In the Appendix~\ref{app:Lepton number conservation}, we prove that this source term does not violate the lepton number conservation.

In our practical calculation, we prepare a table of the following values in advance
\begin{eqnarray}
A^0(\varepsilon,\varepsilon',\mu_N,T)&=&\int d\mu'\delta A_\pm(\varepsilon,\varepsilon',\mu',b)/(\mu_B^N b) \\
A^1(\varepsilon,\varepsilon',\mu_N,T)&=&\int d\mu'\mu'\delta A_\pm(\varepsilon,\varepsilon',\mu',b)/(\mu_B^N b),
\end{eqnarray}
for $\varepsilon(\varepsilon')$, which is the neutrino's energy grid used, and for typical values of chemical potential of free nucleons $\mu_N$ and matter temperature $T$ in CCSNe.
In the above equations, we factored out the magnetic field dependence $\mu_B^N b$ from the tabulated value, which can be incorporated later in the simulation from the local value.
During the simulation, we interpolate the values $A^{0,1}$ along $\mu_N$ and $T$ directions at each energy grid $\varepsilon$ and $\varepsilon'$, multiply them by $\mu_B^N b$, and evaluate Eqs.~(\ref{eq:0thangularmoment_sc_a1}), (\ref{eq:0thangularmoment_sc_c0}),  (\ref{eq:1stangularmoment_sc_c1}), and (\ref{eq:Final4sourceterm_sc_bneq0}).

\subsection{Modified charged current reactions with free nucleons}
\label{sec:Modified charged current reactions with free nucleons}
In this section, we briefly recapitulate how the magnetic field alter the charged current processes ($\nu_e n\rightleftharpoons e^{-} p$ and $\bar\nu_e p\rightleftharpoons e^{+} n$) based on AL99.
In the presence of magnetic field, the energy of electron and positron is quantized, namely Landau quantization.
Due to this quantization, the Fermi-Dirac distribution of final state electron and positron is also affected.
According to AL99, the absorptivity $1/\lambda$ is modified due to the external magnetic field as
\begin{flalign}
\lambda^{-1}=\lambda_0^{-1}(1+\epsilon_{\rm mc} l^\alpha \hat b_\alpha).
\label{eq:ModifiedAbsorptivity}
\end{flalign}
Here $\lambda_0^{-1}$ is the absorptivity without the influence of magnetic field.
$\epsilon_{\rm mc}$ is the correction factor due to magnetic field and expressed as
\begin{flalign}
\epsilon_{\rm mc}=\epsilon_{\rm mc}(e^-)+\epsilon_{\rm mc}(np),
\label{eps_mc_nue}
\end{flalign}
for the reaction $\nu_e n\rightarrow e^- p$, and
\begin{flalign}
\epsilon_{\rm mc}=\epsilon_{\rm mc}(e^+)+\bar\epsilon_{\rm mc}(np),
\label{eps_mc_barnue}
\end{flalign}
for the reaction $\bar\nu_ep\rightarrow e^+ n$.
The correction terms $\epsilon_{\rm mc}(e^-)$ and $\epsilon_{\rm mc}(e^+)$ are originated from electrons and positrons at ground-state Landau level.
While, $\epsilon_{\rm mc}(np)$ and $\bar\epsilon_{\rm mc}(np)$ are from polarized neutrons and protons.

All these terms have a linear dependence on the magnetic field strength $b$ as follows (in cgs units):
\begin{eqnarray}
\label{eq:emc_e}
\epsilon_{\rm mc}(e^-)=\frac{1}{2}\frac{\hbar ceb}{(\varepsilon+Q)^2}\frac{g_V^2-g_A^2}{g_V^2+3g_A^2},
\end{eqnarray}
\begin{eqnarray}
\label{eq:emc_p}
\epsilon_{\rm mc}(e^+)=\frac{1}{2}\frac{\hbar ceb}{(\varepsilon-Q)^2}\frac{g_V^2-g_A^2}{g_V^2+3g_A^2},
\end{eqnarray}
\begin{eqnarray}
\label{eq:emc_np}
\epsilon_{\rm mc}(np)&=&\frac{2g_A(g_A+g_V)}{g_V^2+3g_A^2}\frac{\mu_B^nb}{T}\nonumber\\
&&-\frac{T}{\varepsilon+Q}\left[1+\frac{\varepsilon+Q}{T}f_{e^-}(\varepsilon+Q)\right]\nonumber \\
&&\times\left[\frac{2g_A(g_A+g_V)}{g_V^2+3g_A^2}\frac{\mu_B^nb}{T}
+\frac{2g_A(g_A-g_V)}{g_V^2+3g_A^2}\frac{\mu_B^pb}{T}\right],\nonumber\\
\end{eqnarray}
and
\begin{eqnarray}
\label{eq:baremc_np}
\bar\epsilon_{\rm mc}(np)&=&-\frac{2g_A(g_A-g_V)}{g_V^2+3g_A^2}\frac{\mu_B^pb}{T}\nonumber\\
&&+\frac{T}{\varepsilon-Q}\left[1+\frac{\varepsilon-Q}{T}f_{e^+}(\varepsilon-Q)\right]\nonumber \\
&&\times\left[\frac{2g_A(g_A-g_V)}{g_V^2+3g_A^2}\frac{\mu_B^pb}{T}
+\frac{2g_A(g_A+g_V)}{g_V^2+3g_A^2}\frac{\mu_B^nb}{T}\right].\nonumber\\
\end{eqnarray}
In the above equations, $g_V=1$ and $g_A=1.23$ are the nucleon charged current form factors \citep{Bruenn85}, $f_x(\varepsilon)=[1+\rm{exp}((\varepsilon-\mu_{\it x})/T)]^{-1}$ represents the Fermi-Dirac distribution function of fermion $x$ with energy $\varepsilon$ and chemical potential $\mu_{\it x}$, and $Q=m_n-m_p=1.2935$ MeV is the rest mass difference of the neutron and proton.

From the modified absorptivity $1/\lambda$, the emissivity $j$ can be obtained by
\begin{flalign}
j=\lambda^{-1} \exp((\mu_\nu-\varepsilon)/T),
\label{eq:ModifiedEmissivity}
\end{flalign}
with $\mu_{\nu_e}=-\mu_{\bar\nu_e}=\mu_e-\mu_p+\mu_n$ being the chemical potential of neutrinos in thermal equilibrium with matter.
Then the collision rate for neutrino absorption and emission processes (nae) becomes as
\begin{equation}
B_{\rm nae}(\varepsilon,\Omega)=\kappa\left[f(\varepsilon,\Omega)-f_\varepsilon^{\rm eq}\right],
\label{dfdt_abs}
\end{equation}
with $\kappa=j+\lambda^{-1}$ being the stimulated absorption opacity.

\subsection{Moment formalism of the charged current reaction including magnetic field correction}
\label{sec:Moment formalism of the charged current reaction including magnetic field correction}
Now, analogously to Sec.~\ref{sec:Moment formalism of the modified neutrino-nucleon scattering rate}, we evaluate the charged current source term in moment formalism including the correction term due to the magnetic field.
After performing the angular integral $\int d\Omega$ of the collision rate
\begin{eqnarray}
S^\alpha_{\varepsilon,{\rm nae}}=\varepsilon^3\int d\Omega B_{\rm nae}(\varepsilon,\Omega) (u^\alpha+l^\alpha)
\label{eq:4sourceterm_abs}
\end{eqnarray}
for neutrino absorption and emission processes, it results again in a summation of normal absorption term $S^{\alpha,b=0}_{\varepsilon,{\rm nae}}$ and correction term $S^{\alpha,b\neq0}_{\varepsilon,{\rm nae}}$ as
\begin{flalign}
S^\alpha_{\varepsilon,{\rm nae}}=S^{\alpha,b=0}_{\varepsilon,{\rm nae}}
+S^{\alpha,b\neq0}_{\varepsilon,{\rm nae}}.
\label{eq:Final4sourceterm_nae}
\end{flalign}
Here
\begin{eqnarray}
S^{\alpha,b=0}_{\varepsilon,{\rm nae}}&=&\kappa \left[(J^{\rm eq}-J)u^\alpha-H^\alpha  \right]
\label{eq:Final4sourceterm_nae_beq0}\\
 S^{\alpha,b\neq0}_{\varepsilon,{\rm nae}}&=&\kappa \left[
 -\epsilon_{\rm mc} (H^\beta u^\alpha+\tilde L^{\alpha\beta}) \hat b_\beta \right].
\label{eq:Final4sourceterm_nae_bneq0}
\end{eqnarray}

\section{Initial models and neutrino opacities}
\label{sec:Initial models and neutrino opacities}
Utilizing our $\nu$-GRMHD code including the parity violated source term described above, we perform full 3D CCSN simulations of a magnetized rotating star. 
Numerical setup is essentially the same as that of our previous paper \cite{KurodaT20} other than the neutrino opacity.
We study the frequently used solar-metallicity model of the 20 $M_{\odot}$ star ``s20a28n'' from \cite{WH07}.
For the nuclear EOS, we use SFHo of \cite{SFH}.
The 3D computational domain is a cubic box with $3\times10^4$~km width in which nested boxes with 10 refinement levels are embedded.
Each box contains $64^3$ cells\footnote{In \cite{KurodaT20}, there is a typo in the number of grids. It should be $64^3$ cells that cover each level of nested structure, i.e., the same resolution as in this study.} and the minimum grid size near the origin is $\Delta x=458$m.
The neutrino energy space $\varepsilon$ logarithmically covers from 1 to 300 MeV with 12 energy bins.

We shortly mention the updated neutrino opacities.
In \cite{KurodaT20}, we adopted the standard weak interaction set in \citet{Bruenn85} plus nucleon-nucleon Bremsstrahlung \citep{Hannestad98}.
In this study, however, we use up-to-date neutrino opacities based on \cite{Kotake18} in addition to the correction terms arising from the external magnetic field described in the previous section.
One of differences is that we replace the electron capture rate on heavy nuclei by the most elaborate one following \cite{juoda}.
Furthermore, we also take into account following corrections: inelastic contributions and weak magnetism corrections \citep{horowitz02}, the density-dependent effective nucleon mass \citep{Reddy99}, the quenching of the axial-vector coupling constant \citep{Carter&Prakash02,Fischer16}, many-body and virial corrections \citep{horowitz17}, and strangeness contribution to the axial-vector coupling constant \citep{horowitz02}.
We therefore employ all opacities listed in Table 1 in \cite{Kotake18} with excluding set3 and set4.
The source term can now be summarized as
\begin{eqnarray}
S^\alpha_\varepsilon&=&S^{\alpha,b=0}_{\varepsilon}+S^{\alpha,b\neq0}_{\varepsilon},
\label{eq:Source_final}
\end{eqnarray}
with
\begin{eqnarray}
S^{\alpha,b=0}_{\varepsilon}&=&S^{\alpha,b=0}_{\varepsilon,{\rm nae}}+S^{\alpha,b=0}_{\varepsilon,{\rm sc}}+S^{\alpha}_{\varepsilon,{\rm nes}}+S^{\alpha}_{\varepsilon,{\rm tp}}+S^{\alpha}_{\varepsilon,{\rm brem}} \nonumber \\
\label{eq:Source_final_b=0}\\
\label{eq:Source_final_b/=0}
S^{\alpha,b\neq0}_{\varepsilon}&=&S^{\alpha,b\neq0}_{\varepsilon,{\rm nae}}
+S^{\alpha,b\neq0}_{\varepsilon,{\rm sc}}.
\end{eqnarray}
Here, the subindices ``nes'', ``tp'', and ``brem'' stand for the neutrino-electron inelastic scattering, thermal neutrino pair production and annihilation, and nucleon-nucleon bremsstrahlung, respectively.
We note that this paper does not focus on measuring the impact of these up-to-date neutrino opacities currently used, i.e., Eq.~(\ref{eq:Source_final_b=0}), which will be reported elsewhere \citep[see][for a detailed 1D comparison]{Kotake18}.

The original progenitor model ``s20a28n'' assumes neither rotation nor magnetic field during its evolution phase.
We thus artificially add them to the nonrotating progenitor model as follows:
\begin{equation}
    u^tu_\phi=\varpi_0^2(\Omega_0-\Omega),
\end{equation}
for the rotational profile, and
\begin{eqnarray}
A_\phi&=&\frac{B_0}{2}\frac{R_0^3}{r^3+R_0^3}r\sin{\theta},\\
A_r&=&A_\theta=0,
\end{eqnarray}
for the magnetic field in the form of vector potential.
Here $u_\phi\equiv\sqrt{u_x^2+u_y^2}$ and
$\varpi_0$ and $\Omega_0$ indicate the size and angular frequency of a rigidly rotating central cylinder, respectively.
$B_0$ and $R_0$ represent the magnetic field strength at center and the size of central sphere with uniform magnetic field, respectively.
$(r,\theta,\phi)$ denote the usual coordinates in the spherical polar coordinate system.
By defining the vector potential on the numerical cell edge and taking their curl $\bf B=\nabla\times{\bf A}$, the magnetic field defined on the numerical cell surface automatically satisfies the solenoidal constraint.
These rotational and magnetic field profiles are identical to those used in our former paper \cite{KurodaT20}.

We set $\varpi_0=R_0=10^8$~cm corresponding roughly to the iron core size at the precollapse stage.
We calculate three models with changing $(\Omega_0 [{\rm rad~s}^{-1}],B_0[{\rm G}])$ as $(0,0)$, $(1,10^{12})$, and $(1,10^{13})$, hereafter labeled as R0B00, R1B12, and R1B13, respectively.
The angular frequency $\Omega_0=1$ rad s$^{-1}$ is very reasonable compared to the one of a rotating 20 $M_{\odot}$ model in \cite{Heger00} that gives $\Omega_0\sim3$ rad s$^{-1}$.
Regarding the magnetic field strength inside the iron core at pre-collapse stage, there is currently no constraint from observational and stellar evolution calculation sides.
The value $B_0=10^{12}$~G in model R1B12 is widely used in most of previous MRE simulations \citep{Burrows07,Takiwaki09,Scheidegger10,Moesta14,Obergaulinger20,Bugli20,KurodaT20} and can be a reference case.
On the other hand, $B_0=10^{13}$~G is enormously strong and might be unrealistic. We, however, calculate such an ultra-strongly magnetized model to measure the impact of magnetic field on the neutrino-matter interactions more easily, since the source terms Eq.~(\ref{eq:Final4sourceterm_sc_bneq0}) and (\ref{eq:Final4sourceterm_nae_bneq0}) have a linear dependence on the magnetic field strength $b$.

\section{Results}
\label{sec:Results}
In this section, we present the results of our simulations.
We begin by a detailed explanation of explosion dynamics of our fiducial model R1B12 and then compare three models including neutrino profiles.
Afterward we discuss our main results on the actual impact of magnetic field on the neutrino matter interaction and its possible effects on the dynamics.
As we will show, we see global asymmetric features in the neutrino-matter interactions appearing in the PNS that are closely correlated with the magnetic field structure.
Therefore, we consider that understanding the magnetohydrodynamics inside the PNS is essential to explain those asymmetric features.

\subsection{Explosion dynamics}
\label{sec:Explosion dynamics}
\begin{figure*}[htbp]
\begin{center}
\includegraphics[width=80mm]{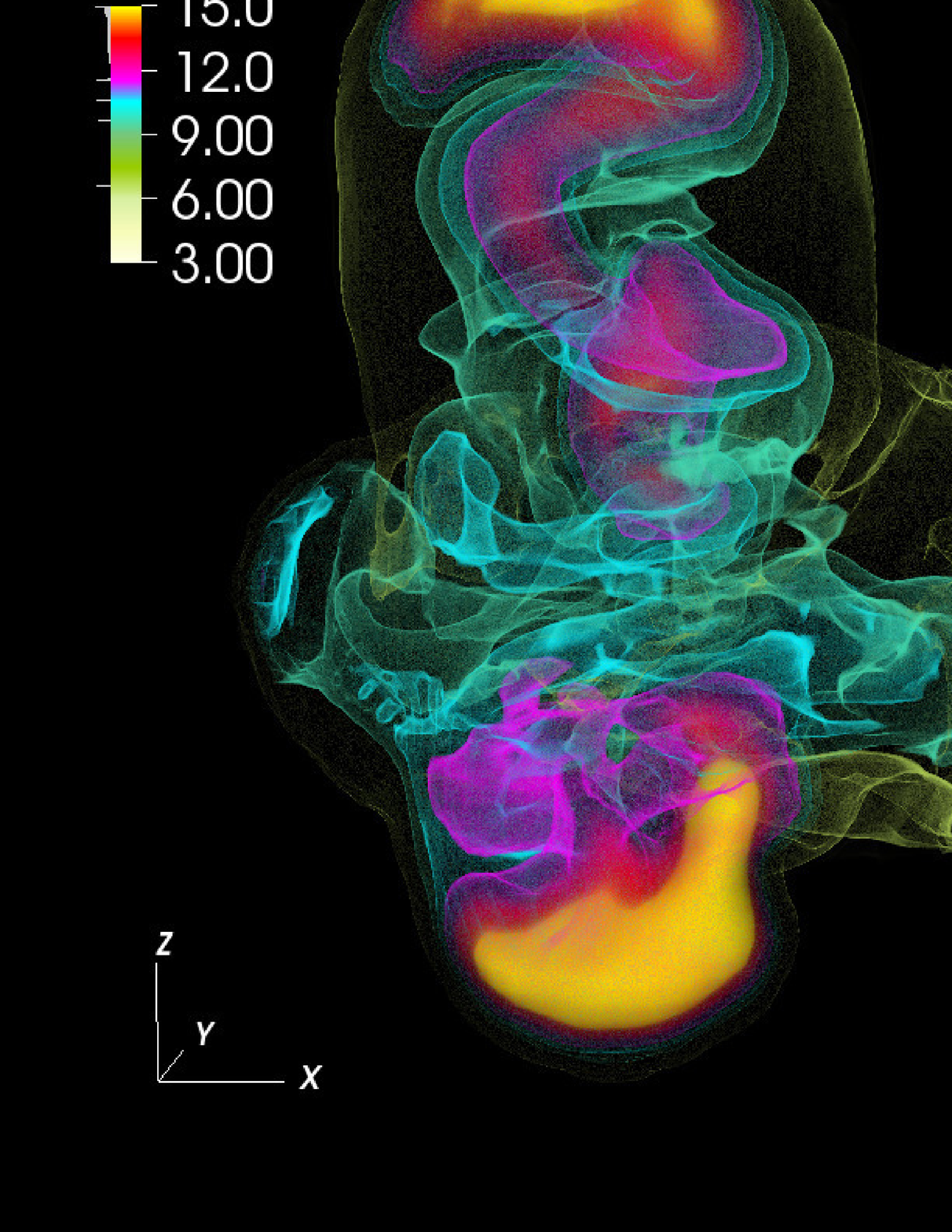}
\includegraphics[width=80mm]{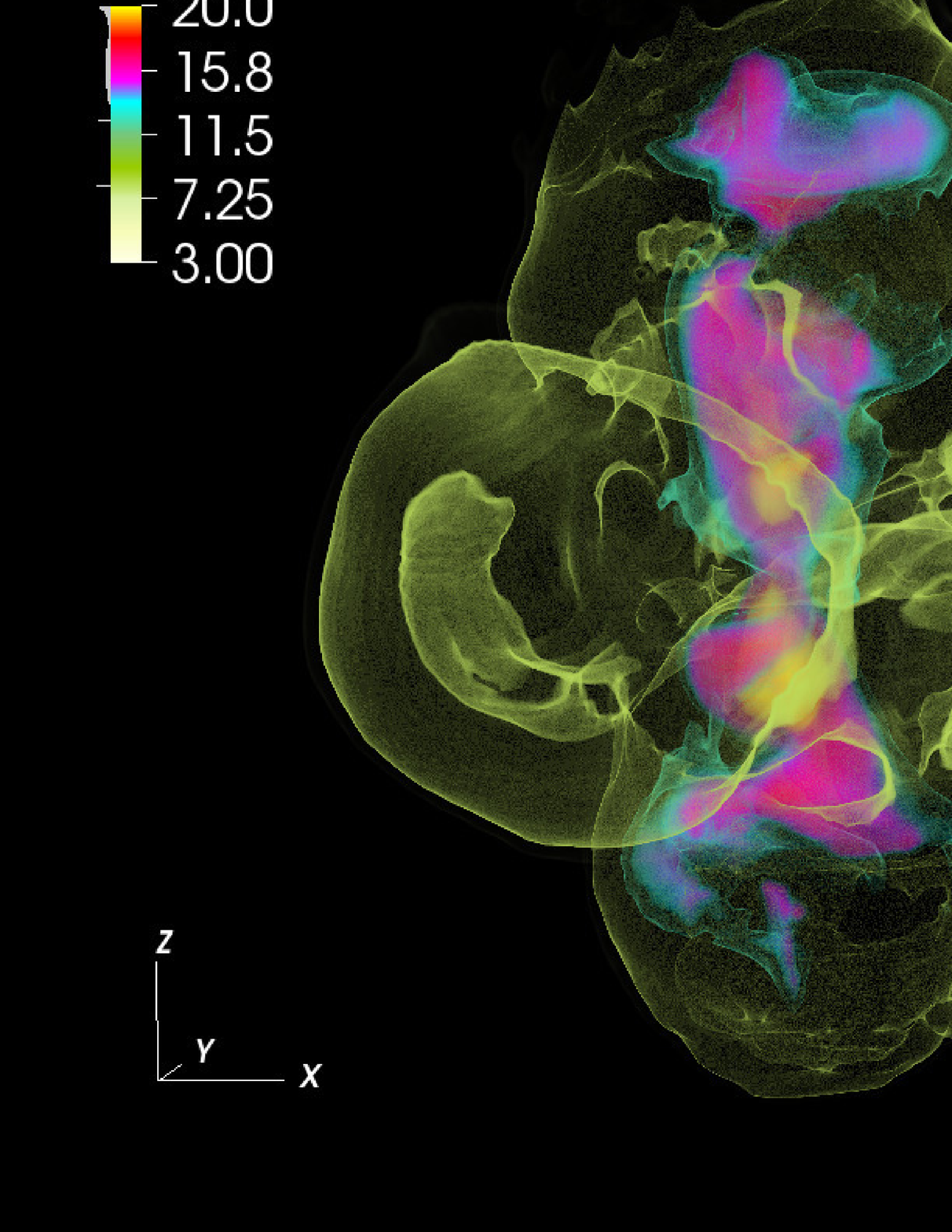}\\
\includegraphics[width=80mm,angle=-90.]{R1B12_Ent.eps}
\includegraphics[width=80mm,angle=-90.]{R1B12_Beta.eps}
  \caption{Top: the volume-rendered entropy for model R1B12 at two different time slices $t_{\rm pb}=183$~ms (left panel) and 367~ms (right).
  The white vertical line indicates the length scale and is also parallel to the rotational axis ($z$-axis).
  Note that the entropy range differs in each panel.
  Bottom: we depict 2D contours of entropy (left) and plasma $\beta$ (right) in logarithmic scale at the final simulation time of $t_{\rm pb}=316$~ms.
  There are three minipanels in the bottom panels.
  Each minipanel shows a 2D slice on $y=0$ plane (minipanel a), $x=0$ (b), and $z=0$ (c).
  \label{fig:3D_2D_R1B12}
}
\end{center}
\end{figure*}

We explain first the dynamical evolution of our fiducial model R1B12.
Its evolution is essentially the same as our previous study employing the same initial condition \citep[model R1B12 in][]{KurodaT20} and the bipolar outflow is launched soon after bounce.
Figure~\ref{fig:3D_2D_R1B12} shows the volume-rendered 3D entropy structure for model R1B12 at two different time slices $t_{\rm pb}=183$~ms (top-left panel) and 367~ms (top-right), where $t_{\rm pb}$ represents the postbounce time, and 2D contours of entropy (bottom-left) and plasma $\beta(\equiv P_{\rm mag}/P_{\rm gas}$), which is the ratio of magnetic to gas pressure, in logarithmic scale (bottom-right) at the final simulation time.
The white vertical line in the top panels indicates the length scale and is also parallel to the rotational axis ($z$-axis).
There are three minipanels in the bottom two panels.
Each minipanel shows a 2D slice on $y=0$ (minipanel a), $x=0$ (b), and $z=0$ (c) planes.
From the figure, we see a clear bipolar shock structure which continuously expands without a stall.
The size of the shock surface increases from $\sim1000$~km to $\gtrsim4000$~km for the time interval of $\sim180$~ms.
Model R1B12 is thus considered to be entering the shock runaway phase directly after bounce.

Its explosion morphology exhibits a clear bipolar-like structure with a slight asymmetry with respect to the equatorial plane.
From bottom minipanels (a,b), the expansion toward the positive $z$-axis is more energetic than the negative direction.
The bipolar structure consists of the high entropy ejecta and the entropy increases with time.
The forefront of the bipolar jet shows the highest entropy of $s\sim15$ $k_{\rm B}$ baryon$^{-1}$ at $t_{\rm pb}\sim183$~ms, while the vicinity of the base of jets shows the highest value exceeding $s\sim20$ $k_{\rm B}$ baryon$^{-1}$ at $t_{\rm pb}\sim367$~ms (yellowish region in the top-right panel).
Furthermore, the jet barycenter shows a clear displacement from the rotational axis, like a helical structure seen in the top-left panel, indicating the appearance of the kink instability \cite[e.g.,][]{Begelman98}.
This is consistent with the previous full 3D MHD CCSN simulations \citep{Moesta14,KurodaT20}.
From the bottom-right panel, we see that the magnetic pressure dominates over the gas pressure inside the outflow (red region), indicating that the MR-driven explosion occurs in this model.

Regarding the north-south asymmetry seen in the explosion morphology, one might expect that it could be caused by the parity violation effects.
We, however, consider that the asymmetry appearing during our simulation time, i.e., the early postbounce phase of $t_{\rm pb}\lesssim$ a few 100 ms, is unlikely due to the parity violation effects of weak interactions, but mostly due to the MHD effects.
In our previous study \citep{KurodaT20}, which did not take into account the parity violation effects, the north-south asymmetry was also found.
In the literature, we introduced a combination of the $m=1$ rotational instability and the MHD kink instability as one of possible explanations of the north-south asymmetry.
By comparing the degree of the asymmetry between this and former studies, we recognize that there is not a significant difference, especially in the initial postbounce phase of $t_{\rm pb}\lesssim100$ ms during which the parity violation effects are particularly strong (explained in the next subsection \ref{sec:North-south asymmetry of neutrino matter interaction}).
More precisely, the corresponding normalized mode amplitudes of spherical polar expansion of the shock surface $A_{\ell m}$ read $\sim0.05$ \% in both studies for ($\ell, m$)=(1,0).
We thus consider that the parity violation effects play a subdominant role, at least in the early postbounce phase, in forming the shock morphology.

On the equatorial plane, a clear $m=1$ nonaxisymmetric structure becomes prominent (see the minipanels c).
Along the equatorial plane, both a continuous mass accretion, with low entropy $s\lesssim5$ $k_{\rm B}$ baryon$^{-1}$, and ejection, with relatively high $s\sim10$ $k_{\rm B}$ baryon$^{-1}$, simultaneously take place.
Furthermore, from minipanel (c) in the bottom-right panel, we find the gas pressure being the dominant in most of the regions, i.e., the magnetic field do not play the leading role for the shock expansion.
We thus argue that the shock expansion is significantly aided by rotation in the equatorial region (see \cite{Nakamura3D14,takiwaki16,Summa18} for the rotation-supported 3D CCSN models and also \cite{KurodaT20} with magnetic field).

We also calculate a supplement model R1B13, which adopts one order of magnitude larger initial magnetic field, to focus on its possibly more emphasized parity violation effects.
Thus, we leave its detailed explanation of (magneto)hydrodynamics in Appendix~\ref{app:Explosion dynamics of a supplement model R1B13}.
Roughly speaking, this model also exhibits a very energetic shock expansion immediately after bounce.
However, the magnetic field structure as well as the entropy structure within the shocked region are significantly different from those in model R1B12.
They do not show a simple bipolar-like structure compared to those in model R1B12, but a slightly complicated structure.

\begin{figure}[htbp]
\begin{center}
\includegraphics[width=75mm,angle=-90.]{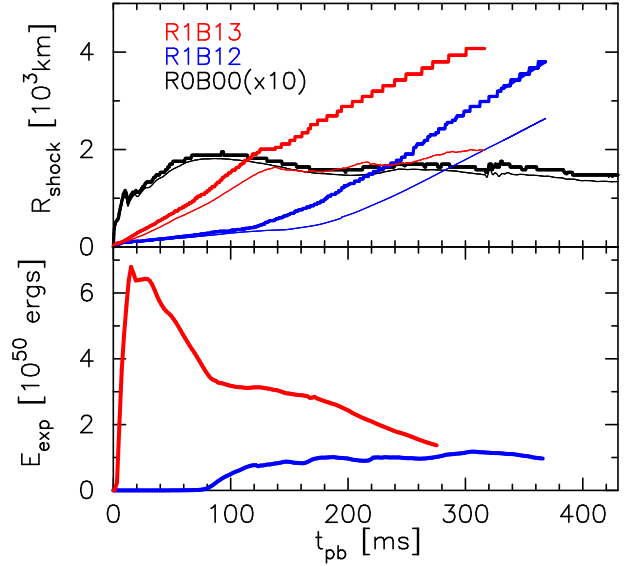}
  \caption{Top: time evolution of maximum (thick lines) and averaged shock radii (thin) for all models. To show more clearly the lines of model R0B00, we multiply them by ten.
  Bottom: we plot the diagnostic explosion energy.
  Since $E_{\rm exp}$ of model R0B00 is essentially zero, we omit to plot it.
  \label{fig:RshockEexp}
}
\end{center}
\end{figure}
Figure~\ref{fig:RshockEexp} presents time evolution of the shock radius $R_{\rm shock}$ in the top panel and of the diagnostic explosion energy $E_{\rm exp}$ in the bottom, where we use the same definition for $E_{\rm exp}$ as \cite{KurodaT20}.
In the top panel, we plot the maximum (thick line) and averaged shock radii (thin) for all models.
We multiply the lines of $R_{\rm shock}$ for model R0B00 (black lines) by ten to show them more clearly.
We find that the model R0B00, which assumes neither rotation nor magnetic field at initial, does not exhibit the shock revival during our simulation time of $\sim500$~ms after bounce.
This is consistent with our previous study \cite{KurodaT20}.
We thus argue that the up-to-date neutrino opacities do not drastically change the explosion dynamics.
Therefore, in the bottom panel, we omit the line for model R0B00 because its explosion energy is essentially zero.

The time evolution of shock radii presents a rapidly expanding shock surface in model R1B13 immediately after bounce (red lines).
The maximum shock radius reaches $R=1000$~km at $t_{\rm pb}\sim70$~ms.
During the same period, $E_{\rm exp}$ increases drastically at initial, reaches its maximum $\sim6\times10^{50}$ ergs at $t_{\rm pb}\sim20$~ms, and declines afterward.
Contribution of each energy to the total diagnostic explosion energy at $t_{\rm pb}=20$ ms is as follows: the magnetic energy $\sim10^{51}$ ergs, the internal energy $\sim4\times10^{50}$ ergs, the radial kinetic energy $\sim2\times10^{50}$ ergs, and the rotational kinetic energy $\sim2\times10^{50}$ ergs.
The prompt explosion is thus mainly supported by the magnetic field.
The decline seen in $E_{\rm exp}$ after $t_{\rm pb}\sim20$ ms is mostly due to decrease of the magnetic energy in the unbound region.
While in model R1B12 (blue lines), the shock front shows a mild expansion at initial ($t_{\rm pb}\lesssim120$ ms) and subsequently becomes faster than that of model R1B13.
Its averaged shock radius (thin blue line) exceeds that of model R1B13 at $t_{\rm pb}\sim310$ ms.
The explosion energy $E_{\rm exp}$ in model R1B12 increases at $t_{\rm pb}\sim100$~ms and then plateaus around the value of $\sim10^{50}$ ergs.
The shock and explosion energy evolution in models R1B12 and R0B00 are quantitatively in good agreement with our previous report \citep{KurodaT20}, which employed the same initial condition except the neutrino opacity set.

\begin{figure}[htbp]
\begin{center}
\includegraphics[width=50mm,angle=-90.]{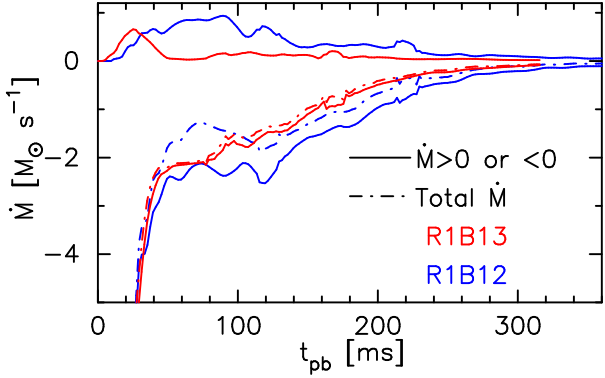}
  \caption{We plot the mass inflow ($\dot M<0$) and outflow rate ($\dot M>0$) (solid lines) and total accretion rate (dash-dotted) measured at $R=100$~km for models R1B13 (red) and R1B12 (blue).
  \label{fig:MdotR100km}
}
\end{center}
\end{figure}
We next explain the different explosion dynamics seen in the two magnetized models, particularly focusing on the reason why the initially less magnetized model R1B12 eventually shows more energetic explosion.
Figure~\ref{fig:MdotR100km} exhibits the mass inflow ($\dot M<0$) and outflow rate ($\dot M>0$) by the solid lines and the total mass accretion rate by the dash-dotted line.
The color represents model either R1B13 (red) or R1B12 (blue).
The rate is measured at $R=100$~km so that we can encompass the base of MHD outflow locating at $R\sim $ a few 10~km.
In model R1B12 (blue lines), the absolute values of both the mass inflow and outflow rates are significantly larger than those of model R1B13 (red).
Especially after $t_{\rm pb}\gtrsim40$~ms, the mass inflow rate in model R1B12 is showing a larger value of $\sim1\ \dot M_{\odot}\ s^{-1}$ from the one in R1B13.
At the same time, the large amount of mass ejection also takes place in model R1B12 resulting in a moderate difference between the total mass accretion rates of two models (dash-dotted lines).
The smaller mass inflow rate in model R1B13 is a consequence of the stronger explosion taking place immediately after bounce.
As we have already explained, enormously strong shock wave, which is mainly supported by the magnetic pressure, is launched shortly after bounce.
The shock propagates outward in all directions and suppresses the subsequent mass accretion onto the central engine of the MHD outflow.
Lower accretion produces less liberation gravitational potential energy and thus the strong bipolar flow is not continuously supported.
This is the reason of the weaker shock wave and also different magnetic field and entropy configurations eventually seen in model R1B13.

On the other hand, the shock propagates significantly slower in model R1B12 at initial postbounce phase, that allows the development of nonaxisymmetric matter motion, e.g., one-armed spiral pattern.
Indeed, the ratio of rotational to gravitational potential energy reaches a few \% in model R1B12, while it is significantly smaller $\sim0.5$ \% in R1B13.
The spiral pattern then produces a flow channel through which noticeable amount of mass accretion takes place as we have shown in the bottom panels of Fig.~\ref{fig:3D_2D_R1B12}.


\subsection{North-south asymmetry of neutrino matter interaction}
\label{sec:North-south asymmetry of neutrino matter interaction}
In this section, we discuss the modified neutrino-matter interactions in the presence of magnetic field and their actual impact on the dynamics.
We begin with a brief explanation of overall picture of neutrino signals.
The nonexplosion model R0B00 shows basically the highest luminosity and mean energy in all flavors of neutrinos, while the rotating magnetized models R1B12 and R1B13, which explode shortly after bounce, show rather lower values.
Roughly speaking, such a feature is consistent with our previous study \citep{KurodaT20} and also with those in recent studies with detailed neutrino transport \citep{BMuller17,O'Connor18,Summa18,Vartanyan19}.

\begin{figure*}[htbp]
\begin{center}
\includegraphics[angle=-90.,width=160mm]{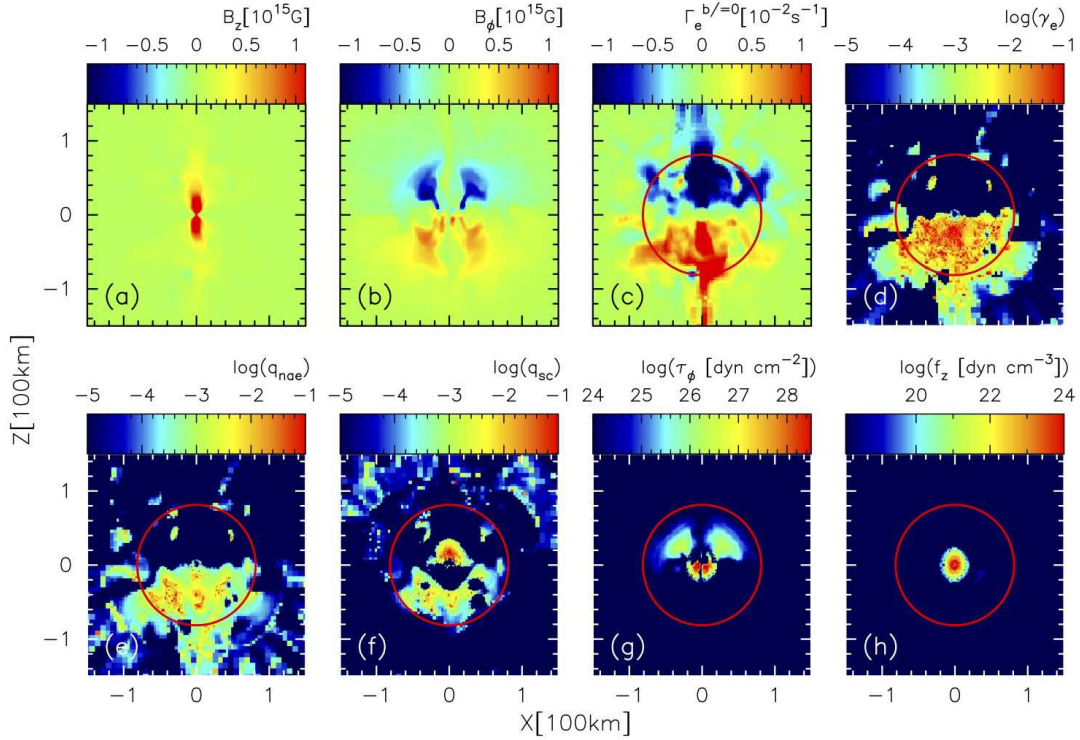}
  \caption{We show: (a) the $z$-component of magnetic field $B_z$ in units of $10^{15}$~G, (b) the $\phi$-component of magnetic field $B_\phi$ in units of $10^{15}$~G, (c) $\Gamma_e^{b\neq0}$ in units of s$^{-1}$, (d) $\log\gamma_e$, (e) $\log q_{\rm nae}$, (f) $\log q_{\rm sc}$, (g) the torque per volume that the fluid element gains $\tau_\phi$ dyn~cm$^{-2}$ in logarithmic scale, and (h) the force acting on the fluid element per volume $f_z$ dyn~cm$^{-3}$ in logarithmic scale.
  Concerning the values in panels (d)-(h), where the values are shown in logarithmic scale, we do not take their absolute values and the negative values are simply cutoff at the minimum value indicated by dark blue.
  The figure is for model R1B12 at $t_{\rm pb}=65$~ms on $y=0$ plane.
  \label{fig:R1B12_STY000600}
}
\end{center}
\end{figure*}

To see more precisely how the modified neutrino-matter interactions lead to asymmetric properties, particularly with respect to the equatorial plane, we depict several quantities in Fig.~\ref{fig:R1B12_STY000600}.
Here, we introduce several quantities as follows:
the change rate of electron fraction $\Gamma_e(\equiv\partial Y_e/\partial t) $ independent from $\left(\Gamma_e^{b=0}\right)$ and dependent on $\left(\Gamma_e^{b\neq0}\right)$ the magnetic field
\begin{eqnarray}
\Gamma_e^{b=0/b\neq0}&=&\frac{\alpha\sqrt{\gamma}}{n_{\rm b}}\int \frac{d\varepsilon}{\varepsilon}S^{\mu,b=0/b\neq0}_\varepsilon u_\mu,
\end{eqnarray}
where $n_{\rm b}$ is the number density of baryons, the energy deposition rate $Q^{b\neq0}_{\rm nae}$ originated from the modified neutrino absorption and emission process
\begin{eqnarray}
Q^{b\neq0}_{\rm nae}&=&\alpha\sqrt{\gamma}\int d\varepsilon S^{\mu,b\neq0}_{\varepsilon, {\rm nae}} n_\mu,
\end{eqnarray}
the energy deposition rate $Q^{b\neq0}_{\rm sc}$ originated from the modified inelastic neutrino-nucleon scattering process
\begin{eqnarray}
Q^{b\neq0}_{\rm sc}&=&\alpha\sqrt{\gamma}\int d\varepsilon S^{\mu,b\neq0}_{\varepsilon, {\rm sc}} n_\mu,
\end{eqnarray}
the energy deposition rate $Q_{\rm tot}$ originated from the normal neutrino matter interactions (see Eq.~\ref{eq:Source_final_b=0})
\begin{eqnarray}
Q_{\rm tot}&=&\alpha\sqrt{\gamma}\int d\varepsilon S^{\mu,b=0}_\varepsilon n_\mu,
\end{eqnarray}
the torque $\tau_\phi$ that the fluid element gains due to the modified source term $S^{\alpha,b\neq0}_{\varepsilon}$
\begin{eqnarray}
\tau_\phi&=&-\alpha\sqrt{\gamma}\varpi \int d\varepsilon S^{\mu,b\neq0}_\varepsilon \gamma_{y\mu},
\end{eqnarray}
where $\varpi=\sqrt{x^2+y^2}$ is the distance from the rotational axis, and the $z$-component of the force acting on the fluid element $f_z$ due to the modified source term $S^{\alpha,b\neq0}_{\varepsilon}$
\begin{eqnarray}
f_z&=&-\alpha\sqrt{\gamma}\int d\varepsilon S^{\mu,b\neq0}_\varepsilon \gamma_{z\mu}.
\label{eq:fz}
\end{eqnarray}
We also introduce following ratios to assess the relative impact of the modified neutrino matter interactions to the normal ones as
\begin{eqnarray}
\gamma_e&=&\frac{\Gamma_e^{b\neq0}}{|\Gamma_e^{b=0}|}\\
q_{\rm nae/sc}&=&\frac{Q^{b\neq0}_{\rm nae/sc}}{|Q_{\rm tot}|}.
\end{eqnarray}
We note that we take an absolute value of the denominator for the later convenience.
Using these quantities, we show in Fig.~\ref{fig:R1B12_STY000600}: (minipanel a) the $z$-component of magnetic field $B_z$, (b) the $\phi$-component of magnetic field $B_\phi$, (c) $\Gamma_e^{b\neq0}$, (d) $\log\gamma_e$, (e) $\log q_{\rm nae}$, (f) $\log q_{\rm sc}$, (g) $\log\tau_\phi$, and (h) $\log f_z$.
All panels are showing the contours on $y=0$ plane for model R1B12 at $t_{\rm pb}=65$~ms.
We also plot the mean gain radius by a red circle in panels (c)-(h) for a reference.

In panel (a), we see that $B_z$ is positive inside the PNS with its value being $\sim10^{15}$~G in both hemispheres.
This is mostly due to compression of initial uniform or dipole-like magnetic field aligning with the $z$-axis inside $\varpi\lesssim1000$~km.
$B_\phi$ shows a clear asymmetry with respect to the equatorial plane and the northern ($z>0$) and southern ($z<0$) hemisphere possess negative and positive $B_\phi$, respectively.
Within the PNS, the toroidal magnetic field strength reaches $\sim10^{15}$~G.
Reflecting the nonisotropic magnetic field, the change rate of electron fraction $\Gamma_e^{b\ne0}$ also shows a clear north-south asymmetry in panel (c).
Here we note that $\Gamma_e^{b\ne0}$ is affected only by the modified neutrino absorption and emission process (Eq.~\ref{eq:Final4sourceterm_nae_bneq0}), since the (modified) scattering process Eq.~(\ref{eq:Final4sourceterm_sc_bneq0}) does not change the electron number at all as proved in the Appendix~\ref{app:Lepton number conservation}.
The panel (c) shows that the northern hemisphere delptonizes more.
This can be understood from Eq.~(\ref{eq:Final4sourceterm_nae_bneq0}).
Using the orthogonality condition $\tilde L^{\alpha\beta}u_\alpha=0$, $\Gamma_e^{b\ne0}$ is expressed as
\begin{eqnarray}
\Gamma_e^{b\neq0}&=&\alpha\sqrt{\gamma}\int \frac{d\varepsilon}{\varepsilon}S^{\mu,b\neq0}_\varepsilon u_\mu \nonumber \\
&=&\alpha\sqrt{\gamma}\int \frac{d\varepsilon}{\varepsilon}\kappa \epsilon_{\rm mc} H^\beta \hat b_\beta.
\end{eqnarray}
$\epsilon_{\rm mc}$ (Eq.~\ref{eps_mc_nue}) is usually negative due to $g^2_V-g^2_A<0$ and also in the degenerate limit of electrons $f_{e^-}\sim1$.
Therefore, the sign of $\Gamma_e^{b\neq0}$ is determined by the angle between the direction of flux of neutrinos $H^\beta$ and that of magnetic field $\hat b_\beta$.
Since $H^\beta$ of $\nu_e$ is usually orienting radially outward inside the PNS, i.e., $H^i\sim (H^r,0,0)$, the toroidal magnetic field $B_\phi$ does not contribute to $\Gamma_e^{b\neq0}$ as they are mutually orthogonal.
The remaining magnetic field component $B_z$ is pointing positive $z$ direction from panel (a).
As a consequence, the northern hemisphere, where $H^\beta \hat b_\beta>0$, $\Gamma_e^{b\neq0}$ becomes negative, while the southern hemisphere analogously shows positive $\Gamma_e^{b\neq0}$.

Panel (d) indicates how large is the contribution of $\Gamma_e^{b\neq0}$ to the total deleptonization rate $|\Gamma_e^{b=0}|$.
Here, we note that the most of the northern hemisphere displays dark blue region simply because we cutoff the negative value to emphasize the north-south asymmetry.
However, it actually has a similar value as in the southern hemisphere, but with a different sign.
We also mention that the deleptonization rate $\Gamma_e^{b=0}$, which is free from the influence of magnetic field, shows a nearly perfect symmetry with respect to the equatorial plane.
From panel (d), the value $\gamma_e$ in the southern hemisphere reaches several \%, which is also the same for the northern hemisphere.
The parity violation in weak interaction due to the external magnetic field thus can potentially produces a significant partial distribution of $Y_e$ in the PNS, that will also be discussed later.
As $\Gamma_e^{b\neq0}$ in panel (c) shows $\sim\pm0.01$ s$^{-1}$ along the rotational axis, if the parity violation effect lasts $\sim100$~ms, the cumulative change reaches $\delta Y_e\sim\pm10^{-3}$.
In previous studies, \cite{Tamborra14ApJ} reported the existence of the lepton number emission self-sustained asymmetry (LESA) (see also, e.g., \cite{O'Connor18,Powell19,Vartanyan19}), which is originated from a partial distribution of $Y_e$ in the PNS convection zone ($r\sim25$~km).
In their subsequent paper \citep{Glas19}, they explained the origin of the partial distribution of $Y_e$, dominated mainly by the $l=1$ mode, by the PNS convection.
Aside from the PNS convection, our study shows for the first time that the neutrino matter interactions in the presence of strong external magnetic field can be another possible mechanism to produce a partial distribution of $Y_e$.

Next we evaluate the impact of heating and cooling rates contributed from the modified charged and neutral current reactions in panels (e) and (f), respectively.
Comparing the panels (d) and (e), they show the similar pattern as expected, i.e., $Q^{b\neq0}_{\rm nae}$ has basically the same sign as $\Gamma_e^{b\neq0}$.
The panel (e) indicates the larger heating and cooling rate in the southern and northern hemisphere, respectively.
Employing essentially the same initial magnetic field but with various strengths, \cite{Kotake05} also reported the excess of neutrino heating in the southern hemisphere.
Our result is thus qualitatively consistent with theirs.
In addition, \cite{Kotake05} shows a relative contribution of $Q^{b\neq0}_{\rm nae}$ to the total heating/cooling rate $Q_{\rm tot}$ of the order of $\sim$0.1 \%, which is also in good agreement with ours.
The excess/reduction of the energy exchange rate above the gain radius is several 0.1 \% from panel (e).

Regarding $Q_{\rm sc}^{b\neq0}$ (or $q_{\rm sc}$), $Q_{\rm sc}^{b\neq0}$ again shows a clear asymmetric feature in panel (f).
Inside the central region with $r\lesssim30$~km, the northern and southern hemisphere shows the excess of neutrino heating and cooling, respectively.
Meanwhile, the sign inverts around $r\sim40$~km.
The reason of the asymmetry and sign inversion can be understood as follows.
From Eq.~(\ref{eq:Final4sourceterm_sc_bneq0}) and also using the fact that $a_{1,\alpha}$ is proportional to $b_\alpha$ (see Eq.~\ref{eq:0thangularmoment_sc_a1}), we find that $S_{\varepsilon,\rm{sc}}^{\alpha,b\neq0}$ is consisted of two parts.
One is proportional to $H^\beta b_\beta$ and the other is to $|\tilde L|\sim J-J^{\rm eq}$, namely the deviation of neutrino distribution function from the thermal equilibrium, as
\begin{eqnarray}
S^{\alpha,b\neq0}_{\rm sc}\propto H^\beta b_\beta u^\alpha+\frac{J-J^{\rm eq}}{3}b^\alpha.
\label{eq:Final4sourceterm_sc_bneq0_simple}
\end{eqnarray}
Here we use the relation $h^{\alpha\beta}b_\beta=b^\alpha$.
The source term for the zeroth radiation moment  $Q_{\rm sc}^{b\neq0}$ is then rewritten by taking the norm with $n_\alpha$ as
\begin{eqnarray}
Q_{\rm sc}^{b\neq0}&\sim& \int d\varepsilon S^{\alpha,b\neq0}_{\rm sc}n_\alpha\nonumber \\
&\propto& \int d\varepsilon \left( -WH^\beta b_\beta-\frac{J-J^{\rm eq}}{3}B^iu_i\right).
\label{eq:Final4sourceterm_sc_bneq0_simple_0th}
\end{eqnarray}
Inside the deep PNS core, the radiation field and matter velocity exhibit nearly the isotropic structure, i.e., $l=0$ with $l$ denoting the degree of spherical harmonics, while the magnetic field shows a uniform (dipole like) field for our initial condition, i.e., $l=1$.
As a consequence, both the first and second terms, which are depend on the angle between the magnetic field and $H^\beta$ and $u_i$, respectively, basically change their signs between the northern and southern hemispheres.
This is the reason of the asymmetric structure with respect to the equatorial plane.
The sign inversion seen at $r\sim40$~km depends simply on which term of the integrand becomes the dominant term and cannot be inferred a priori.
It is also noteworthy that $Q^{b\neq0}_{\rm sc}$ has a comparable or even a slightly larger value than $Q^{b\neq0}_{\rm nae}$.
This fact indicates the importance of the modified neutrino-nucleon scattering term due to the magnetic field that is often omitted in the previous studies on the parity violation effects in CCSNe \citep{Kotake05,Suwa14,Dobrynina20}.

We also discuss the parity violation effects on the angular and linear momentum transfer.
In panel (g), we show the torque that the PNS gains.
It is obvious that the PNS core $r\lesssim30$~km is subjected to torsional stress and the northern and southern hemisphere spins down and up, respectively.
A possibility of such torsional effect has been already discussed in \cite{Maruyama14}, in which a region where the direction of rotation is the same with that of magnetic field shows a spin deceleration.
Our result thus supports their discussion.
However, since they considered only the effect of absorption process influenced by the magnetic field, our result with both the scattering and absorption processes considered shows a more complicated spin deceleration or acceleration profile.
Finally from panel (h), we find that the linear momentum transfer occurs mostly in the PNS core.
At there, $f_z$ shows a positive value in both hemispheres without asymmetry with respect to the equatorial plane.
It indicates that the whole PNS core can be kicked toward north at this time.
The symmetric structure seen in $f_z$ can be understood like below.
Here we mention that the main contribution to the momentum transfer comes from the scattering term $S_{\varepsilon,\rm{sc}}^{\alpha,b\neq0}$ and we thus focus on this term.
Using Eq.~(\ref{eq:Final4sourceterm_sc_bneq0_simple}), $f_z$ can be expressed as
\begin{eqnarray}
f_z&\sim& -\int d\varepsilon S^{\alpha,b\neq0}_{\rm sc}\gamma_{z\alpha}\nonumber \\
&\propto& -\int d\varepsilon \left( H^z b_z u_z+\frac{J-J^{\rm eq}}{3}b_z\right),
\label{eq:asymptotic_f_z}
\end{eqnarray}
where we take only the $z$ component for $\alpha$ when we move from the first line to the second.
This is because the dominant magnetic field component is $B_z\sim b_z(>0)$ around the rotational axis in the PNS core.
Bearing in mind that the radiation field and matter are roughly isotropic inside the PNS core, $H^z u_z$ and $J-J^{\rm eq}$ basically show the same sign in both hemispheres.
Therefore, $f_z$ is proportional to $b_z$ and it results in a bulk acceleration of PNS core.
Analogously, the angular momentum exchange has a dependence on the direction of toroidal magnetic field.
Consequently, the anti-parallel toroidal magnetic field between northern and southern hemispheres seen in our magnetized models lead to the spin-down and -up, respectively.
We mention that the sign of $f_z$ can change with time, as we will show later, probably depending on the neutrino profiles $H^\mu$ as well as on $u^\mu$ in Eq.~(\ref{eq:asymptotic_f_z}).

\begin{figure*}[htbp]
\begin{center}
\includegraphics[angle=-90.,width=160mm]{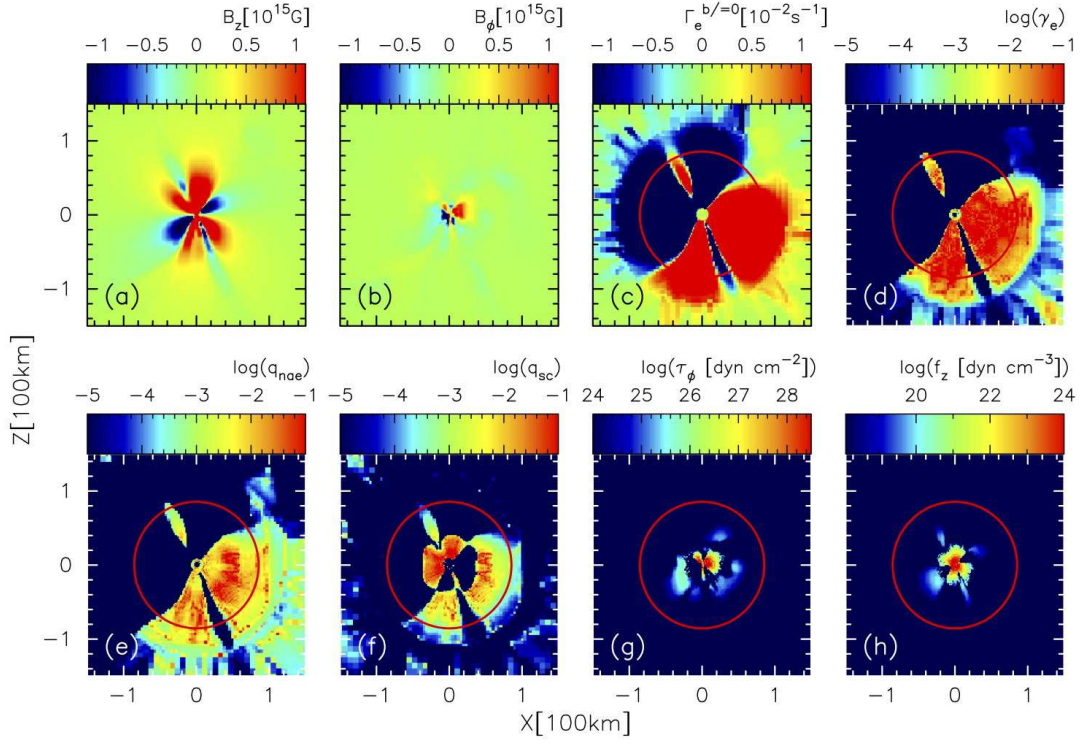}
  \caption{Same as Fig.~\ref{fig:R1B12_STY000600}, but for model R1B13 at $t_{\rm pb}=65$~ms.
  \label{fig:R1B13_STY000715}
}
\end{center}
\end{figure*}
In Fig.~\ref{fig:R1B13_STY000715}, we depict the same values as in Fig.~\ref{fig:R1B12_STY000600}, but for model R1B13 to see the outcomes of one-order of magnitude stronger initial magnetic field.
From panels (a) and (b), the magnetic field are not showing a similar structure as those in model R1B12.
For instance, $B_z$ seems to be consisted of higher order $l$ modes.
As already discussed in Sec.~\ref{sec:Explosion dynamics} and also will be shown in Fig.~\ref{fig:3D_2D_R1B13} in the Appendix~\ref{app:Explosion dynamics of a supplement model R1B13}, the less mass inflow along the equatorial plane might hinders the collimation effect of outflow at where it is launched.
Consequently, the magnetic field at the base of outflow may not be aligned with the rotational axis compared to that in R1B12.
Aside from such difference seen in the magnetic field structure, the effects of modified neutrino matter interactions are qualitatively the same as in R1B12.
Reflecting the global magnetic field, whose direction is roughly tilted with the angle of $\sim-45^\circ$ from the rotational axis on $y$=0 plane (see panel a), the region with $z\gtrsim x$ shows, e.g., the excess of deleptonization and cooling rate, which is analogous to what we see in the northern hemisphere of model R1B12.
The relative contribution of modified neutrino matter interactions to the total ones is approximately several to $\sim10$ \% from panels (d)-(f), which is approximately one order of magnitude larger than that in model R1B12.
This model with different magnetic field structure than model R1B12 clearly shows the importance of magnetic field structure, which potentially determines the global pattern of parity violation effect, and, thus, of magnetohydrodynamic evolution in the PNS.

\begin{figure*}[htbp]
\begin{center}
\includegraphics[angle=-90.,width=170mm]{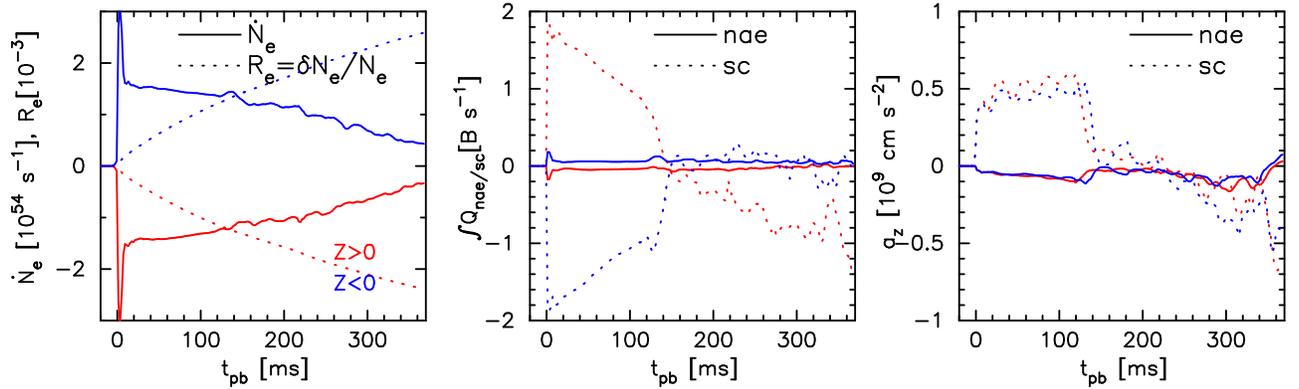}
  \caption{We show the deleptonization rate (left panel), energy deposition rate (middle), and acceleration of PNS $a_z$ (right) in each hemisphere. See text for their definitions.
  In the middle and right panels, the line style corresponds to the neutrino matter interaction, either modified ``nae'' (solid) or ``sc'' (dashed) process.
  \label{fig:PV_R1B12}
}
\end{center}
\end{figure*}
In the end of this section, we discuss time evolution of the modified source terms and their cumulative impact on the PNS core.
In Fig.~\ref{fig:PV_R1B12}, we show the deleptonization rate (left panel), energy deposition rate (middle), and acceleration of PNS (right).
To plot the left panel, we again introduce several quantities as follows: the total deleptonization rate due to the modified interactions
\begin{eqnarray}
\dot N_e&=&\int dx^3 n_{\rm b}\Gamma^{b\neq0}_e
\end{eqnarray}
and its cumulative value
\begin{eqnarray}
\delta N_e&=&\int_{-\infty}^{t} dt \dot N_e,
\end{eqnarray}
number of electrons
\begin{eqnarray}
N_e&=&\int dx^3 \rho^\ast Y_e,
\end{eqnarray}
and the ratio $R_e$, which measures the degree of partial distribution of electrons inside the PNS,
\begin{eqnarray}
R_e&=&\frac{\delta N_e}{N_e}.
\end{eqnarray}
The volume integral is performed for each hemisphere $z>0$ (northern hemisphere, red lines) and $z<0$ (southern hemisphere, blue) and also for the region with $\rho\ge10^{12}$~g~cm$^{-3}$.
In the middle panel, the energy gain or loss in each hemisphere is evaluated by the volume integral of $Q_{\rm nae/sc}^{b\neq0}$.
In the right panel, we evaluate the $z$ component of PNS acceleration $a_z$ by
\begin{eqnarray}
a_z=\frac{1}{M_{\rm PNS}}\int dx^3 f_z,
\label{eq:az}
\end{eqnarray}
where $M_{\rm PNS}=\int dx^3 \rho^\ast$ is the PNS mass in each hemisphere. 
In both the middle and right panels, the solid and dashed lines indicate the contribution from modified neutrino absorption and emission (nae) and scattering (sc) processes, respectively.

From the left panel in Fig.~\ref{fig:PV_R1B12}, the volume integrated deleptonization rate in each hemisphere exhibits a clear asymmetry throughout the simulation time.
$\dot N_e$ in the northern hemisphere reaches $\dot N_e\sim-10^{54}$ s$^{-1}$ (red solid line) indicating more deleptonization at there.
Its cumulative value reaches $R_e\sim-0.2$ \% of the total electron number inside the PNS at the final time of simulation.
Meanwhile, in the southern hemisphere, those values have the opposite sign, but with almost the same absolute values as those in the northern hemisphere.
Therefore, the parity violation effect can potentially produce a north-south asymmetry in $Y_e$ of the order of a few \textperthousand~ in this model.
Regarding the energy deposition rate in the middle panel, significant energy transfer occurs mainly during the first hundred milliseconds after bounce.
However, from panels (e) and (f) in Fig.~\ref{fig:R1B12_STY000600}, the energy transfer due to these terms takes place mostly within the gain radius denoted by the red circle.
We thus argue that the modified terms do not have a noticeable impact on the neutrino heating explosion mechanism.

The apparent PNS acceleration $a_z$, that is simply evaluated from the source term of momentum exchange, shows a significant value of $\sim5\times10^8$~cm~s$^{-2}$ and lasts $\sim100$~ms after bounce.
We note that the momentum exchange through the normal neutrino matter interaction processes without influence of magnetic field, i.e., via the term $S_\varepsilon^{\alpha,b=0}$ (Eq.~\ref{eq:Source_final_b=0}), shows a nearly, not perfect, asymmetric property with respect to the equatorial plane.
For instance at $t_{\rm pb}=50$~ms, $a_z$ evaluated from $S_\varepsilon^{\alpha,b=0}$ reaches $\sim+7.1(-7.3)\times10^{10}$~cm~s$^{-2}$ in the northern(southern) hemisphere, resulting in a non-vanishing net acceleration of $\sim-2\times10^{9}$~cm~s$^{-2}$.
We note that our nonrotating nonmagnetized model R0B00 shows the net acceleration of $\sim10^{7}$~cm~s$^{-2}$, which is essentially zero.
We thus consider that the non-vanishing acceleration contributed solely from $S_\varepsilon^{\alpha,b=0}$ originates from the asymmetric hydrodynamic background with respect to the equatorial plane.
Anyway, we argue that the symmetric property seen in $a_z$ evaluated from $S_\varepsilon^{\alpha,b\neq0}$ (right panel in Fig.~\ref{fig:PV_R1B12}) has a comparable influence as that from $S_\varepsilon^{\alpha,b=0}$.
We also mention that we do not observe a meaningful PNS core kick during the simulation time.
This is due to that the relevant neutrinos are still trapped and do not carry away the momentum during the simulation time.
It marginally shows a slight oscillation after $t_{\rm pb}\sim100$~ms with the displacement of several hundred meters, possibly indicating the appearance of SASI.
Another remarkable thing is that the modified scattering term is again the main contribution by comparing the dotted (sc) and solid (nae) lines in the middle and right panels.

\begin{figure*}[htbp]
\begin{center}
\includegraphics[angle=-90.,width=170mm]{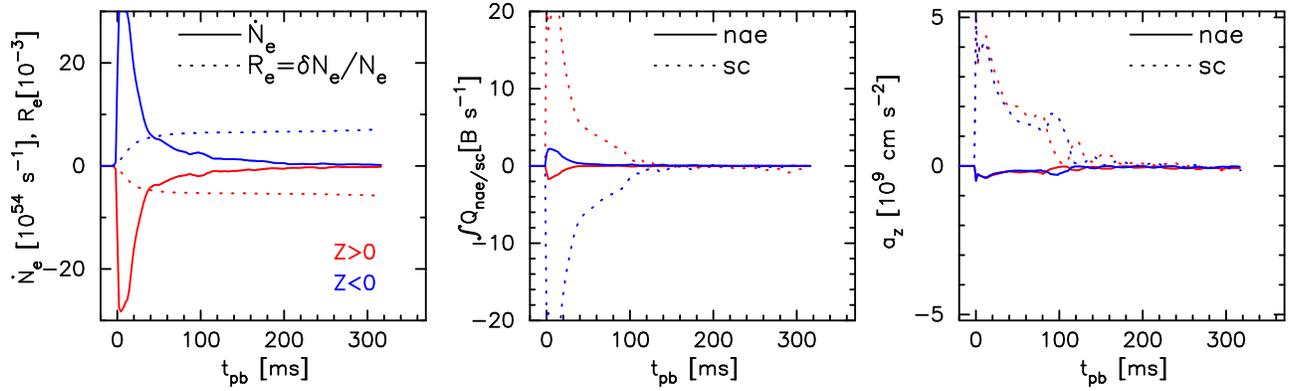}
  \caption{Same as Fig.~\ref{fig:PV_R1B12}, but for model R1B13.
  \label{fig:PV_R1B13}
}
\end{center}
\end{figure*}
In Fig.~\ref{fig:PV_R1B13}, we plot the same figure as Fig.~\ref{fig:PV_R1B12}, but for model R1B13 which employs one order of magnitude larger initial magnetic field.
We see a qualitatively similar trend as that of R1B12.
During the first $\sim100$~ms after bounce, most of the lepton, energy, and momentum exchanges occur.
However, compared to the values in model R1B12, all values plotted in Fig.~\ref{fig:PV_R1B13} show roughly one order of magnitude larger values.
This is simply because that the modified source terms employed (Eqs.~\ref{eq:Final4sourceterm_sc_bneq0} and
\ref{eq:Final4sourceterm_nae_bneq0}) have a linear dependence on the magnetic field $b_\mu$.

\section{Summary and Discussion}
\label{sec:Summary and Discussion}
In this study, we have presented a formalization of the correction term of neutrino matter interaction rates in the presence of external magnetic field.
The formalism is based on \cite{Arras&Lai99} and we took into account the modified interaction rates of two major processes in the SN core: neutrino-nucleon scattering and neutrino absorption and emission processes.
We extracted the zeroth and first order angular dependencies of the interaction rates and derived the source term suitable for the M1 neutrino transport in full relativity.
The final expression of the source terms is described in terms of the normal radiation moments.
Because the magnetic potential energy of free electrons and nucleons, $\hbar ceb$ and $\mu_B b$, respectively, are significantly smaller than the matter temperature in typical MHD CCSN models, we can safely truncate the second or higher order terms in the magnetic field strength, leading to the source terms having a linear dependence on it.
We also proved that the modified scattering term does not violate the lepton number conservation, which is crucial to accurately follow the PNS deleptonization.

Utilizing state-of-the-art general relativistic M1 neutrino transport code with the gravitational red and Doppler shift terms being fully considered, we have conducted MHD CCSN simulations of a 20 $M_\odot$ star \citep{WH07}.
We calculated three models, nonrotating nonmagnetized (R0B00), rotating magnetized (R1B12), and rotating strongly magnetized (R1B13) models to explore the effects of progenitor's rotation and magnetic field both on the dynamics and the modified neutrino matter interactions.
For the nuclear EOS, we used SFHo of \cite{SFH}.
Other than the correction terms in neutrino matter interactions due to the magnetic field, one of major differences from our previous study \citep{KurodaT20} is that we used up-to-date neutrino opacities based on \cite{Kotake18}.
We adopted the most elaborate electron capture rate on heavy nuclei following \cite{juoda}.
Furthermore, we also took into account followings: inelastic contributions and weak magnetism corrections \citep{horowitz02}, the density-dependent effective nucleon mass \citep{Reddy99}, the quenching of the axial-vector coupling constant \citep{Carter&Prakash02,Fischer16}, many-body and virial corrections \citep{horowitz17}, and strangeness contribution to the axial-vector coupling constant \citep{horowitz02}.
Investigating the impact of these up-to-date neutrino opacities will be reported elsewhere.

Concerning the dynamics, while no shock revival was observed in model R0B00 during our simulation time of $\sim500$~ms after bounce, the shock expansion initiated shortly after bounce in two magnetized models.
We found essentially the same dynamical features between models R0B00/R1B12 in this study and their corresponding models in \cite{KurodaT20}.
The only difference is that the bipolar explosion appeared in model R1B12 in this study did not diminish, although it showed a slight asymmetry with respect to the equatorial plane.
In model R1B13, we observed the most rapid shock expansion and largest increment of the explosion energy among the three models.
However the expansion speed of model R1B12 eventually seemed to be faster than that of model R1B13 at the final simulation time.
The explosion energy in model R1B12 reached the value of $\sim10^{50}$ ergs, which is consistent with our previous study.
The reason why the initially less magnetized model R1B12 eventually exhibited more energetic shock expansion is originated from the larger mass accretion rate.
In model R1B12, we witnessed significantly larger mass inflow and outflow rates than those in model R1B13.
Compared to the case in R1B13, the prompt shock propagated significantly slower in model R1B12 that allowed nonaxisymmetric matter motion, e.g., one-armed spiral pattern, to fully develop.
The spiral pattern could then produce a flow channel through which noticeable amount of mass accretion took place, leading to more liberation of gravitational energy.

One of the aims of this study is to self-consistently assess the actual impact of modified neutrino matter interaction rates on the lepton, energy, and momentum exchanges.
In addition, we focused on the global asymmetry that could be induced by the initial dipole-like magnetic field employed in this study.
From our results, we found a clear asymmetric feature in both the deleptonization and energy deposition rates with respect to the equatorial plane.
As for the asymmetric deleptonization rate, the northern(southern) hemisphere loosed electrons more(less).
The energy deposition rate through the modified neutrino absorption and emission process showed basically the same asymmetric feature with that of the deleptonization rate.
These features can be understood by the dependence of corresponding source term on the inner product of the diffusion flux of neutrinos and the magnetic field.
The sign of the inner product changes between the northern and southern hemispheres if the magnetic field is dominated by a dipole (or more precisely odd $l$-modes) structure.
We also demonstrated that a different magnetohydrodynamic evolution produces a different magnetic field structure within the shocked region leading to a different pattern of parity violation effects.

The cumulative impact of asymmetric deleptonization rate asymptotically reaches $\sim0.1-1$ \%.
Our result also indicates the importance of the modified inelastic scattering process, at least in the explosion phase, that has been often omitted in previous literature \citep{Kotake05,Suwa14,Dobrynina20}.
The energy deposition rate from the modified scattering term showed roughly a one order of magnitude larger value than that of absorption process.

On the contrary to the lepton and energy exchanges, the linear momentum exchange showed a symmetric property, e.g., the $z$ component of the PNS acceleration showed the same sign and value in both hemispheres.
Roughly speaking, such feature stems from that the source term projected on to the Eulerian frame is parallel to the magnetic field.
Initially dipolar-like magnetic field leads to a magnetic field configuration in the proto-magneter composed of a nearly uniform $z$ component and anti-parallel toroidal field in both hemispheres.
Therefore, both hemispheres gain the acceleration in the same $z$ direction, while there is a torsional effect between the two hemispheres.

In terms of scattering and absorption cross sections, we can summarize our results as follows: the scattering cross sections for neutrinos are enhanced(reduced) if neutrinos propagate in parallel(antiparallel) with magnetic field, which is consistent with the positive acceleration of PNS core.
On the contrary, the absorption cross sections are reduced(enhanced) if neutrinos propagate in parallel(antiparallel) with magnetic field, which is in line with, for instance, the smaller deleptonization rate, i.e., neutrinos are less captured, displayed in the northern hemisphere (see panel (c) in Fig.~\ref{fig:R1B12_STY000600}).
These trends of enhancement or reduction are consistent with the study of \cite{Maruyama11,Maruyama12}.

We stress that most of the energy, leptons, and linear and angular momentums that are transferred through the modified source terms are not immediately taken away from deep inside the PNS core.
We actually investigated if there is a noticeable north-south asymmetry in $Y_e$ or neutrino's energy and number fluxes on the PNS surface, though we could not find them with significance.
Therefore we conclude that the modified source terms do not play important roles in the dynamical time scale and contribute neither the explosion dynamics nor the {\it natal} ($t_{\rm pb}\lesssim1$ s) PNS kick.
This basically agrees with previous studies \cite{Kotake05,Dobrynina20}.
However, once those trapped neutrinos diffuse out the PNS core on a time scale of $\gtrsim\mathcal O(1)$ s, namely in the Kelvin-Helmholtz cooling phase, they can potentially produce a $\sim0.1-1$ \% asymmetry in the PNS structure.
Along with the neutrino diffusion, the partial distribution of $Y_e$ may gradually appear leading to more pronounced asymmetric neutrino emission.
In addition, the modified scattering and absorption processes themselves contribute to globally asymmetric neutrino emission.
The final outcome of the asymmetric neutrino signal depends on these two effects that definitely interact each other.
Therefore, the assessment of the degree of final outcome is highly complex and cannot be simply extrapolated from our results.
It can be explored only through the long time MHD simulation.
Our numerical simulations, however, provide us, for the first time, realistic values of the impact of magnetic field on the neutrino matter interactions in MRE scenario, albeit in the early post bounce phase and for a small number of models.

\acknowledgements{
It is a pleasure to thank Almudena Arcones for her encouragement and support for this work and also for her useful comments and suggestions.
I acknowledge Kei Kotake and Tomoya Takiwaki for fruitful discussions and the new neutrino opacities that they provided me.
This research was supported by the ERC Starting Grant EUROPIUM-677912.
Numerical computations were carried out on Cray XC50 at CfCA, NAOJ.
}

\appendix
\section{Lepton number conservation}
\label{app:Lepton number conservation}
It is informative to show that the correction term $S^{\alpha,b\neq0}_{\rm sc}(\varepsilon)$ does not violate the lepton number conservation, i.e., the number of neutrinos does not change through the scattering process.
This is equivalent to satisfy the following condition (see Eq.~(15) in \cite{KurodaT16}):
\begin{eqnarray}
\int \frac{d\varepsilon}{\varepsilon}S^{\alpha,b\neq0}_{\varepsilon,\rm sc} u_\alpha =0
\label{eq:SU=0}
\end{eqnarray}
for every flavor of neutrino.
As we mentioned, the source term Eq.~(\ref{eq:4sourceterm_sc_b=0_bneq0}) consists of the zeroth and first order angular moments.
Regarding the number integral of the first order moment of the source term, i.e., inserting Eq.~(\ref{eq:1stangularmoment_sc}) in Eq.~(\ref{eq:SU=0}), it is zero by definition due to the orthogonality conditions $L^{\alpha\beta}u_\alpha=0$ and $h^{\alpha\beta}u_\alpha=0$.
We mention that the scattering source term with no contribution from magnetic field  Eq.~(\ref{eq:4sourceterm_sc_b=0}) analogously conserves the lepton number because of the condition $H^{\alpha}u_\alpha=0$.

Concerning the remaining zeroth order term Eq.~(\ref{eq:0thangularmoment_sc}), its number integral also becomes zero.
To show that, we first rewrite Eq.~(\ref{eq:0thangularmoment_sc}).
After some manipulation, it becomes
\begin{flalign}
&\varepsilon^3u^\alpha \int d\Omega B_{\rm sc}^{b\neq0}(\varepsilon,\Omega) =\nonumber \\
& 2\pi u^\alpha\Biggl[
\int d\mu' \int d\varepsilon'\Biggl(
\delta A_+DH_\varepsilon^\beta -\biggl(\frac{\varepsilon}{\varepsilon'}\biggr)
\delta A_+'D'H_{\varepsilon'}^\beta
\Biggr) \nonumber \\
&+\int d\mu' \mu' \int d\varepsilon'\Biggl(
-\delta A_+'C'H_\varepsilon^\beta +\biggl(\frac{\varepsilon}{\varepsilon'}\biggr)
\delta A_+CH_{\varepsilon'}^\beta
\Biggr)
\Biggr]\hat b_\beta
.
\label{eq:0thangularmoment_sc_neu_H}
\end{flalign}
Here, we omit the argument $(\varepsilon,\varepsilon')$ in $C$ and $D$ for simplicity.
In addition, $C'$, $D'$, and $\delta A_+'$ with $'$ denote the value with the incoming neutrino energy $\varepsilon$ and that of outgoing one $\varepsilon'$ are switched, e.g., $\delta A'_+\equiv\delta A_+(\varepsilon',\varepsilon,\mu',b)$.
For $\delta A_+$, this manipulation obviously does not alter the scattered angle $\mu'$ and magnetic field strength $b$.
Furthermore, we also use the following relations:
\begin{eqnarray}
{q_0}'&=&-q_0\\
D'&=&-e^{q_0/T} C\\
\varepsilon'^2\delta A'_+C'&=&-\varepsilon^2\delta A_- D\\
\varepsilon'^2\delta A'_+D'&=&-\varepsilon^2\delta A_- C.
\label{eq:relation_DC}
\end{eqnarray}
Consequently, the number integral of Eq.~(\ref{eq:0thangularmoment_sc_neu_H}) results in
\begin{flalign}
&\int\frac{d\varepsilon}{\varepsilon}u_\alpha \left(\varepsilon^3u^\alpha  \int d\Omega B_{\rm sc}^{b\neq0}(\varepsilon,\Omega)\right) =-2\pi \hat b_\beta\nonumber \\
& \times \Biggl[
\int d\mu' \int d\varepsilon d\varepsilon'\Biggl(
\frac{1}{\varepsilon}\delta A_+DH_\varepsilon^\beta -\frac{1}{\varepsilon'}
\delta A_+'D'H_{\varepsilon'}^\beta
\Biggr) \nonumber \\
&+\int d\mu' \mu' \int d\varepsilon d\varepsilon'\Biggl(
-\frac{1}{\varepsilon}\delta A_+'C'H_\varepsilon^\beta +\frac{1}{\varepsilon'}
\delta A_+CH_{\varepsilon'}^\beta
\Biggr)
\Biggr]\nonumber \\
&=0.
\label{eq:number_integral_0thangularmoment_sc}
\end{flalign}
Therefore the condition Eq.~(\ref{eq:SU=0}) is indeed satisfied for $\nu_e$ and analogously for $\bar\nu_e$, as we only have to replace $\delta A_+$ and $\delta A_+^{'}$ in Eq.~(\ref{eq:number_integral_0thangularmoment_sc}) by $\delta A_-$ and $\delta A_-^{'}$, respectively, for $\bar\nu_e$.

\section{Explosion dynamics of a supplement model R1B13}
\label{app:Explosion dynamics of a supplement model R1B13}

\begin{figure*}[htbp]
\begin{center}
\includegraphics[width=80mm]{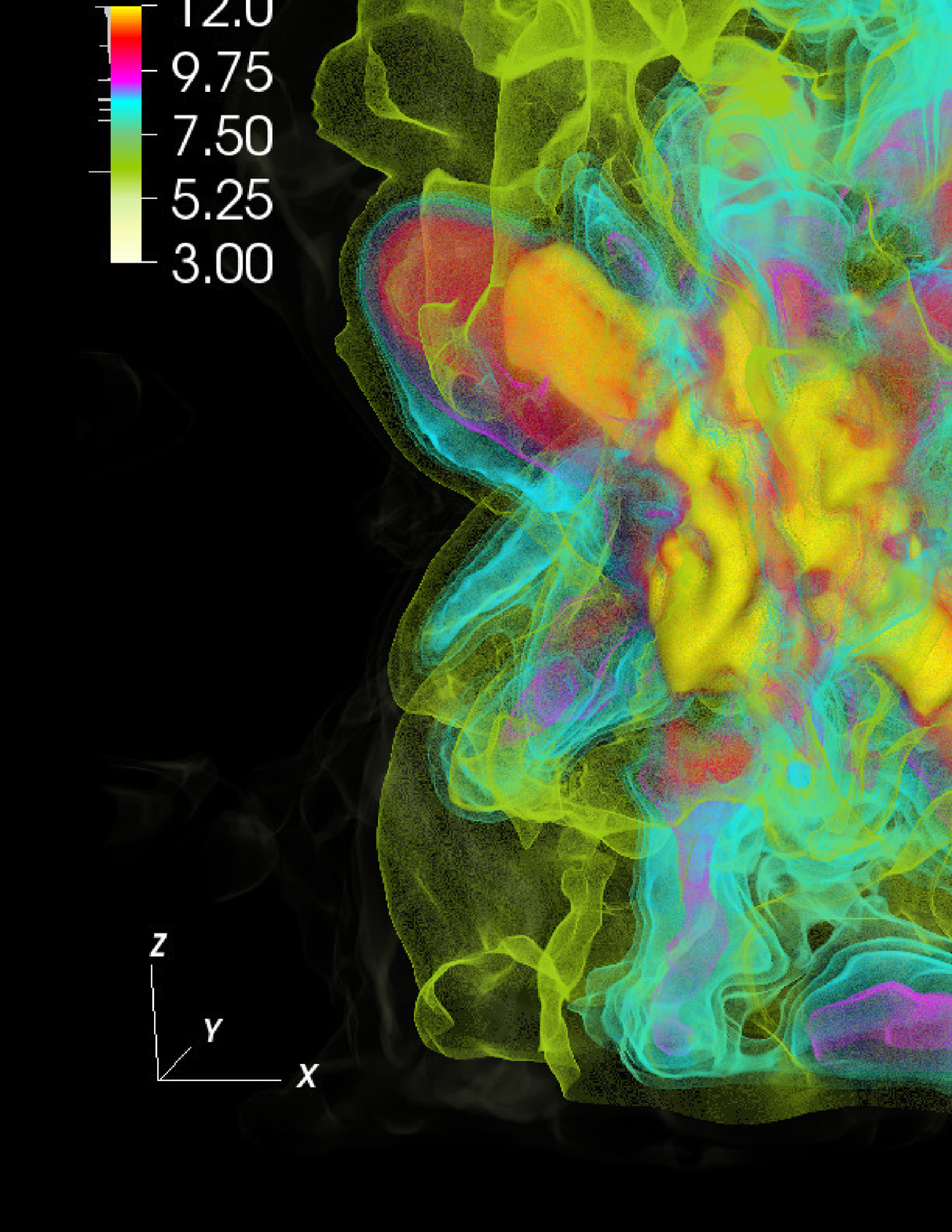}
\includegraphics[width=80mm]{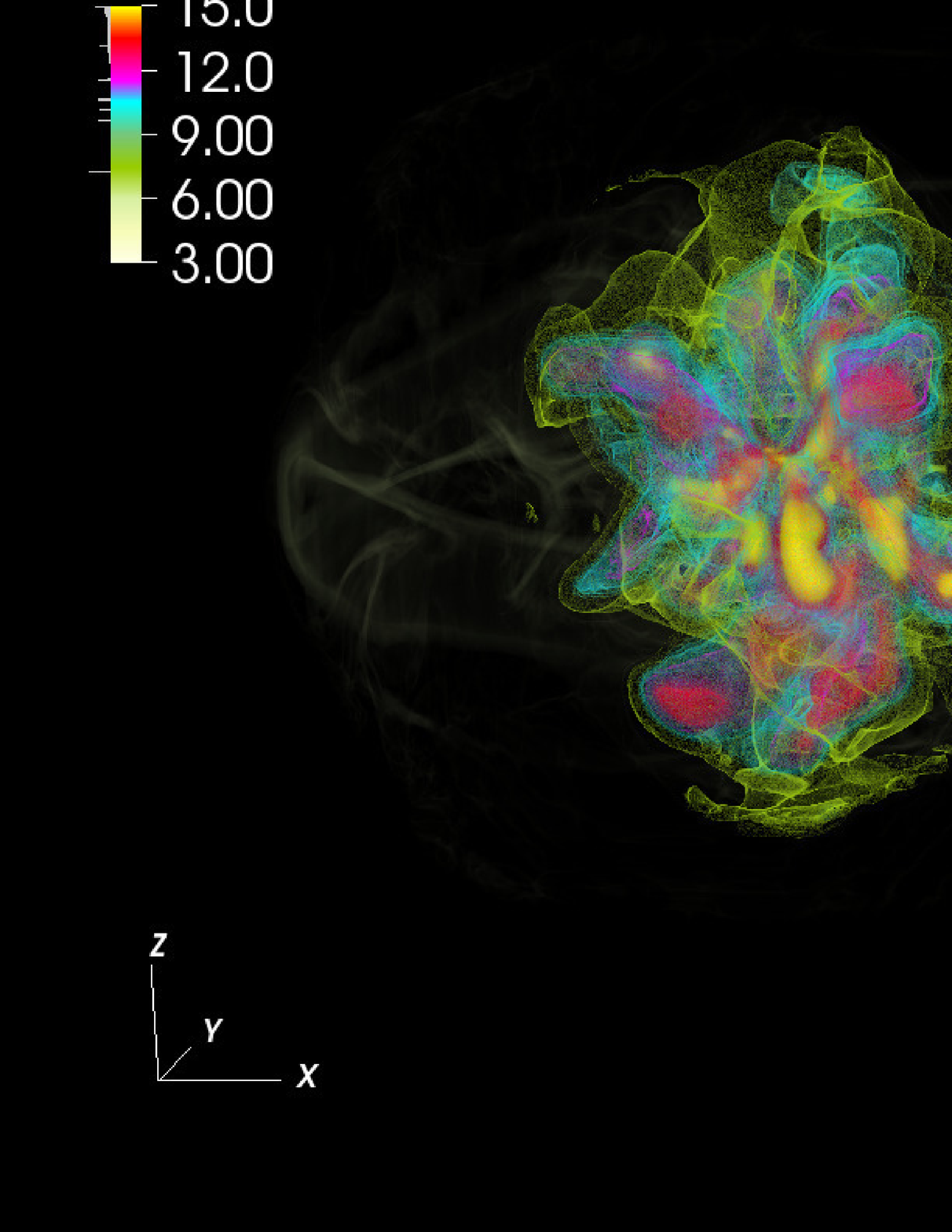}\\
\includegraphics[width=80mm,angle=-90.]{R1B13_Ent.eps}
\includegraphics[width=80mm,angle=-90.]{R1B13_Beta.eps}
  \caption{Same as Figure \ref{fig:3D_2D_R1B12}, but for model R1B13.
  \label{fig:3D_2D_R1B13}
}
\end{center}
\end{figure*}
In this appendix, we briefly explain the explosion dynamics of a supplement model R1B13.
After bounce, this model exhibits approximately one order of magnitude larger magnetic field strength with significantly different structure than a simple dipole-like one seen in model R1B12.
We therefore consider this model as an appropriate model to see how the magnetic field structure determines the global pattern of parity violation effects.

In Fig.~\ref{fig:3D_2D_R1B13}, we show the same figure as Fig.~\ref{fig:3D_2D_R1B12}, but for model R1B13 at $t_{\rm pb}=185$~ms and at the final simulation time $t_{\rm pb}=316$~ms.
The remarkable difference from model R1B12 is the absence of a clear bipolar structure.
For instance from 2D contours in the bottom panels, the outermost shock surface locating at $R\sim4000$~km is more roundish and the axis ratio is closer to unity than that of model R1B12.
Inside the shocked region, the high entropy region with $s\sim15$ $k_{\rm B}$ baryon$^{-1}$ appears, although the value is relatively lower compared to model R1B12 throughout the computation.
In addition, the entropy structure shows small scale and fragmented structure that is completely different from a global scale high entropy bipolar structure seen in model R1B12.
Another remarkable feature is that the high entropy blobs (e.g., yellowish region in the top two panels) are randomly oriented and basically not in alignment with the rotational axis in contrast to model R1B12.
Reflecting such a feature, the magnetic field also exhibits a more stochastic structure, as can be partly understood from the plasma $\beta$ (in the lower right panels of Fig.~\ref{fig:3D_2D_R1B13}), than that of model R1B12, which, in contrast, shows a bipolar structure consisted of high plasma $\beta$ gas.

We consider that the reason of this random orientation is due to the absence of continuous magnetic field amplification, particularly the winding amplification along the rotational axis.
As will be explained later, this model shows a less mass accretion rate and the winding mechanism, which requires high mass accretion rate with angular momentum, does not operate.
Due to lack of the strong toroidal magnetic field along the rotational axis, the matter is not preferentially ejected along the rotational axis.
In addition, there might be another possible reason for the misalignment, which is the tilt of rotational axis.
Although it is not the scope of this work to investigate in detail if the rotational axis tilts in model R1B13, from a very recent full 3D MHD models in \cite{Obergaulinger20_3D}, the rotational axis, as well as the orientation of the outflow, can tilt due to asymmetric matter accretion onto the PNS.
We found that the main mass accretion in model R1B13 takes place not only along the equatorial plane but also in all directions.
It significantly differs from that of model R1B12, which shows a continuous accretion channel mainly along the equatorial plane.
Although \cite{Obergaulinger20_3D} reported the tilt of rotational axis in their weakest magnetized model, the absence of continuous mass accretion along the equatorial plane is a common feature seen in our model R1B13 and theirs.

\bibliographystyle{aasjournal}
\bibliography{mybib}

\end{document}